\documentclass[10pt,aps,prd,floatfix,nofootinbib,superscriptaddress,preprintnumbers]{revtex4-2}
\pdfoutput=1

\usepackage{graphicx} 
\usepackage{amsmath,amssymb,bm}
\usepackage{xcolor}
\usepackage[colorlinks,allcolors=blue]{hyperref} 

\begin{document}

\preprint{TU-1262}

\title{Q-balls Under Spontaneously Broken U(1) Symmetry}

\author{Masahiro Kawasaki}
\email{kawasaki@icrr.u-tokyo.ac.jp}
\affiliation{ICRR, University of Tokyo, Kashiwa 277-8582, Japan}
\affiliation{Kavli IPMU (WPI), UTIAS, University of Tokyo, Kashiwa 277-8583, Japan}
\author{Kai Murai}
\email{kai.murai.e2@tohoku.ac.jp}
\affiliation{Department of Physics, Tohoku University, Sendai, Miyagi 980-8578, Japan}
\author{Fuminobu Takahashi}
\email{fumi@tohoku.ac.jp}
\affiliation{Department of Physics, Tohoku University, Sendai, Miyagi 980-8578, Japan}
\affiliation{Kavli IPMU (WPI), UTIAS, University of Tokyo, Kashiwa 277-8583, Japan}

\begin{abstract}
We study the evolution of Q-balls under a spontaneously broken global $U(1)$ symmetry. Q-balls are stabilized by the conservation of $U(1)$ charge, but when the symmetry is spontaneously broken, the resulting Nambu-Goldstone (NG) boson can carry charge away from the Q-ball, potentially leading to charge leakage. To study this process in a controlled setting, we consider a scenario where Q-balls first form under an unbroken $U(1)$ symmetry, which is then spontaneously broken. We introduce two complex scalar fields: one responsible for forming the Q-ball, and the other for spontaneously breaking the $U(1)$ symmetry, allowing us to clearly separate the formation and symmetry-breaking phases.
Using numerical simulations in a spherically symmetric system, we find that the evolution of Q-balls depends sensitively on the structure of the interaction between the two fields and the magnitude of symmetry breaking. Depending on parameters, Q-balls can completely decay, evaporate into smaller, stable Q-balls, or transition into oscillons/I-balls. 
In particular, we find that stable, localized remnants often survive the evolution over long timescales, especially when the symmetry-breaking scale is small. 
These results demonstrate that, even though spontaneous $U(1)$ breaking can lead to significant energy and charge loss from Q-balls, stable localized objects with reduced or no charge can frequently survive and potentially contribute to cosmological relics.
\end{abstract}

\maketitle

\section{Introduction}
Q-balls~\cite{Coleman:1985ki} are non-topological solitons arising in scalar field theories with a global $U(1)$ symmetry. They are stabilized by a conserved $U(1)$ charge and appear as localized lumps of a complex scalar field, with the charge carried by the coherent phase rotation of the field. Classically, Q-balls are stable as long as the underlying $U(1)$ symmetry is preserved.

However, it is generally believed that any global $U(1)$ symmetry is explicitly broken, for example, by non-renormalizable operators suppressed by a high energy scale such as the Planck scale. This expectation is motivated by arguments from quantum gravity, which suggest that exact global symmetries are incompatible with the principles of black hole physics~\cite{Banks:2010zn}. Therefore, global $U(1)$ symmetries observed in nature are, at best, approximate.

Baryon and lepton numbers are typical examples of such approximate global $U(1)$ symmetries. They are well conserved at low energies, and can support the formation and stability of Q-balls. However, they are expected to be violated at high energies—for instance, through non-renormalizable operators. In the Affleck-Dine (AD) mechanism~\cite{Affleck:1984fy,Dine:1995kz}, one of the flat directions in the minimal supersymmetric standard model (MSSM) carries baryon and/or lepton number and receives an explicit $U(1)$-breaking A-term in the potential. This A-term gives a kick in the phase direction, initiating rotation in field space and generating a net baryon and/or lepton charge.
Depending on the potential shape, the flat direction may experience spatial instabilities to fragment into Q-balls~\cite{Kusenko:1997zq,Kusenko:1997si,Enqvist:1997si,Enqvist:1998en,Kasuya:1999wu,Kasuya:2000wx,Enqvist:2000gq,Hiramatsu:2010dx}.
Since the $U(1)$ breaking A-term becomes more effective when the field amplitude is large, its impact is particularly significant when Q-balls carry large charge or grow in charge after formation. In such cases, the A-term could destabilize Q-balls or hinder the growth of the Q-ball charge~\cite{Kawasaki:2005xc,Kasuya:2014ofa,Cotner:2016dhw,Kawasaki:2019ywz}. The (in)stability of Q-balls affects, for instance, the viability of Q-ball dark matter~\cite{Kusenko:1997si,Dine:2003ax,Enqvist:2003gh} and the constraint on Q-balls by the observations of neutron stars~\cite{Kusenko:1997it,Kusenko:2005du}.
Thus, the stability of Q-balls depends critically on how well the underlying global $U(1)$ symmetry is preserved.

A different class of global $U(1)$ symmetry appears in the Peccei-Quinn (PQ) mechanism~\cite{Peccei:1977hh,Peccei:1977ur}, which offers the most compelling solution to the strong CP problem. In this case, the global $U(1)$ PQ symmetry must be of extremely high quality—that is, any explicit breaking must be highly suppressed—to ensure the effectiveness of the PQ mechanism. The PQ symmetry is spontaneously broken by the vacuum expectation value (VEV) of a complex scalar field $\Psi$, giving rise to the axion as a (pseudo) Nambu-Goldstone (NG) boson~\cite{Preskill:1982cy,Abbott:1982af,Dine:1982ah}.

This raises a central question: \textit{Can Q-balls remain stable when the global $U(1)$ symmetry is spontaneously broken?}
A Q-ball is, by definition, a localized configuration with finite charge and energy, realized by a coherent phase rotation of the scalar field within the Q-ball. The field amplitude vanishes at spatial infinity ensuring that the total energy remains finite. In this setup, there are no light degrees of freedom associated with the phase at large distances. However, if the same complex scalar field also spontaneously breaks the $U(1)$ symmetry, it must acquire a nonzero VEV even far from the Q-ball core. In that regime, the phase direction corresponds to a massless NG mode and remains dynamical at spatial infinity.
In a single-field setup, this means that the rotational degree of freedom associated with the Q-ball’s charge at the core necessarily extends to the region where the field asymptotes to a nonzero VEV. As a result, any stationary configuration would require the phase rotation to persist all the way to spatial infinity, leading to infinite charge and energy. This suggests that a single complex scalar field attempting to form a Q-ball while simultaneously breaking the 
$U(1)$ symmetry cannot yield a stable, localized solution in vacuum. 

To resolve this tension between a localized Q-ball solution in vacuum and the spontaneous breaking of the U(1) symmetry, we consider a scenario where Q-balls first form under an unbroken $U(1)$ symmetry, which is then spontaneously broken.%
\footnote{See also Note Added for the related work~\cite{Kobayashi:2025qao} that accompanies the present paper.} To this end, we introduce two complex scalars $\Phi$ and $\Psi$.
In the absence of interactions between them, the system would have a $U(1)_\Phi \times U(1)_\Psi$ global symmetry.
We introduce an interaction term that explicitly breaks one linear combination of these symmetries, leaving a single global $U(1)$.
In our setup, $\Phi$ first begins oscillating and fragments into Q-balls while the $U(1)$ symmetry is still effectively unbroken.
Subsequently, as the universe expands, $\Psi$ develops a nonzero VEV, spontaneously breaking the remaining $U(1)$.
By separating the Q-ball formation and spontaneous symmetry breaking (SSB) in time, we can examine how Q-balls evolve once the supporting $U(1)$ is spontaneously broken.

In this paper, we study the dynamics of Q-balls under the spontaneously broken $U(1)$ symmetry.
Specifically, we consider a setup where the field $\Phi$ has a mass potential with a logarithmic correction, enabling the existence of Q-ball solutions before symmetry breaking.
The field $\Psi$, on the other hand, has a wine-bottle potential and acquires a nonzero VEV at low energies, breaking the remaining $U(1)$.
An interaction of the form $\propto \Phi^{n_1} \Psi^{n_2} + \mathrm{h.c.}$ transmits the effect of $U(1)$ breaking to the dynamics of $\Phi$ through the VEV of $\Psi$.
Although this resembles the A-term in supersymmetric models, here $\Psi$ is a dynamical field and receives backreaction from $\Phi$. 
This backreaction is crucial since the remaining $U(1)$ symmetry is not explicitly broken, and the breaking is purely spontaneous.
To keep the system numerically tractable, we mainly focus on renormalizable interactions, while A-terms in SUSY models are typically higher-dimensional.
We perform spherically symmetric numerical simulations, varying the powers $(n_1, n_2)$ and the VEV of $\Psi$, $\eta$.
We find a general trend that larger $\eta$ makes the Q-ball more unstable.
However, the fate of the unstable Q-ball depends on $(n_1, n_2)$: it can completely decay, shrink into a smaller remnant, or transform into an oscillon/I-ball~\cite{Bogolyubsky:1976yu,Gleiser:1993pt,Kasuya:2002zs} and their variants.
We also find that stable localized objects with reduced or vanishing charge often survive the evolution over long timescales, particularly when the SSB scale $\eta$ is relatively small.
These remnants may remain in the later universe, potentially contributing to cosmological relics such as dark matter.
In addition, scalar radiation and gravitational waves emitted during the Q-ball evolution may leave observable imprints and serve as cosmological probes of this scenario.

The rest of this paper is organized as follows.
In Sec.~\ref{sec: model}, we briefly review the Q-ball solution in a single-field model and explain the two-field model we explore.
In Sec.~\ref{sec: numerical}, we describe our numerical setup and show the results.
The last section is devoted to the summary of our results.

\section{Q-ball and two field model}
\label{sec: model}

First, we briefly review the Q-ball solution in a global $U(1)$ theory with one complex scalar field, $\Phi$.
We consider the Lagrangian:
\begin{align}
    \mathcal{L}
    =
    |\partial_\mu \Phi|^2
    - V_\Phi(|\Phi|)
    \ .
\end{align}
The total energy and charge are given by 
\begin{align}
    E
    &=
    \int \mathrm{d} x^3 \,
    \left(
        |\dot{\Phi}|^2 + |\nabla \Phi|^2 + V_\Phi(|\Phi|) 
    \right)
    \ ,
    \\
    Q 
    &=
    i \int \mathrm{d} x^3 \,
    \left( \dot{\Phi}^* \Phi - \Phi^* \dot{\Phi} \right)
    \ .
\end{align}
The Q-ball solution exists as the lowest energy state for a fixed $U(1)$ charge when the potential satisfies certain conditions given later.
To obtain the Q-ball solution, we introduce a Lagrange multiplier, $\omega$, and minimize the following functional,
\begin{align}
    \mathcal{E}_\omega 
    &=
    E + \omega \left[
        Q - i \int \mathrm{d} x^3 \, \left( \dot{\Phi}^* \Phi - \Phi^* \dot{\Phi} \right)
    \right]
    \nonumber \\
    &=
    \omega Q
    + \int \mathrm{d} x^3 \,
    \left(
        |\dot{\Phi} - i \omega \Phi|^2 + |\nabla \Phi|^2 + V_\Phi(|\Phi|) 
        - \omega^2 |\Phi|^2
    \right)
    \ .
\end{align}
The first term in the integral is minimized by 
\begin{align}
    \Phi(t, \bm{x})
    =
    \frac{1}{\sqrt{2}} \phi(r) e^{i \omega t}
    \ ,
\end{align}
where $r = |\bm{x}|$, and we used the fact that the lowest energy state is spherically symmetric~\cite{Coleman:1985ki}.
We can regard $\phi(r)$ as a real function because the complex phase of $\phi(r)$ should be independent of $r$ after the minimization of the gradient energy.
Then, $\mathcal{E}_\omega$ can be rewritten as 
\begin{align}
    \mathcal{E}_\omega 
    &=
    \omega Q
    + 4\pi \int \mathrm{d} r \, r^2
    \left(
        \frac{1}{2} \left(\partial_r \phi\right)^2 
        + V_\phi(\phi)
        - \frac{1}{2} \omega^2 \phi^2
    \right)
    \ ,
\end{align}
where $V_\phi(\phi) \equiv V_\Phi(\phi/\sqrt{2})$.
The Euler-Lagrange equation becomes
\begin{align}
    \frac{\mathrm{d}^2 \phi}{\mathrm{d} r^2}
    + \frac{2}{r} \frac{\mathrm{d} \phi}{\mathrm{d} r}
    + \omega^2 \phi
    - \frac{\partial V_\phi(\phi)}{\partial \phi}
    =
    0
    \ .
\end{align}
We consider the solution to this equation of motion with the boundary condition  
\begin{align}
    \phi(r\to\infty) = 0
    \ , \quad 
    \partial_r \phi(0) = 0
    \ .
    \label{eq: boundary condition for Q-ball}
\end{align}
This problem can be mapped to the dynamics of a homogeneous scalar field in an expanding universe by interpreting $r$ as cosmic time.
In this analogy, the potential for the scalar field is given by 
\begin{align}
    U(\phi)
    \equiv 
    \frac{1}{2} \omega^2 \phi^2 
    - V_\phi (\phi)
    \ .
\end{align}
The equation of motion admits solutions satisfying the boundary conditions~\eqref{eq: boundary condition for Q-ball} if 
\begin{align}
    \min \left[ \frac{2 V_\phi(\phi)}{\phi^2} \right]
    <
    \omega^2
    <
    \left.
        \frac{\partial^2 V_\phi (\phi)}{\partial \phi^2}
    \right|_{\phi = 0}
    \ .
\end{align}

An example of the potential satisfying this condition is
\begin{align}
    V_\Phi(|\Phi|)
    =
    m^2 \left[ 1 - K \log\frac{|\Phi|^2}{\Phi_c^2} 
    \right] |\Phi|^2
    \ ,
\end{align}
where $m$ is the mass of $\Phi$ at the scale of $\Phi_c$, and $K$ is a real, positive, dimensionless constant.
This type of potential arises in gravity-mediated SUSY breaking scenarios within the MSSM~\cite{Enqvist:1997si,Enqvist:1998en}.
Then, the equation of motion becomes 
\begin{align}
    \frac{\mathrm{d}^2 \phi}{\mathrm{d} r^2}
    + \frac{2}{r} \frac{\mathrm{d} \phi}{\mathrm{d} r}
    + \omega^2 \phi
    - m^2 \left[ 1 - K - K \log\frac{\phi^2}{2 \Phi_c^2} 
    \right] \phi
    =
    0
    \ .
\end{align}
It is known that the solution $\phi(r)$ can be given by a Gaussian ansatz~\cite{Enqvist:1997si}
\begin{align}
    \phi(r) 
    =
    \phi_c e^{-r^2/R^2}
    \ ,
\end{align}
where $\phi_c$ is the value of $\phi(r)$ at $r = 0$, and $R$ is the radial scale of the Q-ball solution.
By substituting the ansatz into the equation of motion, we obtain
\begin{align}
    \frac{4 r^2}{R^4} - \frac{6}{R^2}
    + \omega^2 - m^2 \left[ 1 - K + 2 K \frac{r^2}{R^2} - K \log\frac{\phi_c^2 }{2 \Phi_c^2} \right]
    =
    0
    \ .
\end{align}
Taking $\phi_c = \sqrt{2} \Phi_c$, we obtain 
\begin{align}
    R = \sqrt{\frac{2}{K}} \frac{1}{m}
    \ , \quad 
    \omega = \sqrt{1 + 2K} m
    \ .
\end{align}
Note that this solution does not exist when $K$ is negative.

To incorporate the SSB of the $U(1)$ symmetry, we introduce another complex scalar field, $\Psi$, which has a nonzero VEV at low energies. 
Specifically, we consider the following Lagrangian
\begin{align}
    \mathcal{L}
    =
    |\partial_\mu \Phi|^2 + |\partial_\mu \Psi|^2 
    - V(\Phi, \Psi)
    \ ,
\end{align}
with the potential $V(\Phi,\Psi)$ given by
\begin{align}
    V(\Phi, \Psi)
    =
    m^2 \left[ 1 - K \log\frac{|\Phi|^2}{\Phi_c^2} 
    \right] |\Phi|^2
    +
    \frac{\lambda}{4} \left( |\Psi|^2 - \eta^2 \right)^2
    -
    \left( \kappa \Phi^{n_1} \Psi^{n_2} + \mathrm{h.c.} \right)
    \ .
\end{align}
Here, $\eta$ is the VEV of $|\Psi|$, $n_1$ and $n_2$ are positive integers, and $\lambda$ and $\kappa$ are real positive constants.
Although $\kappa$ can be complex in general, it can be made real by redefining the phase of $\Phi$ or $\Psi$.

In the following, we investigate the Q-ball dynamics in the presence of the coupling between $\Phi$ and $\Psi$, which explicitly breaks one linear combination of two $U(1)$ symmetries associated with $\Phi$ and $\Psi$.
The orthogonal combination of the $U(1)$ symmetries remains unbroken, and it can be identified with the PQ symmetry.
Under the remaining $U(1)$ symmetry, we assign charges $n_2$ and $-n_1$ to $\Phi$ and $\Psi$,  respectively.%
\footnote{The charges could be divided by the common divisor of $n_1$ and $n_2$.}
Although the nonzero VEV at low energies, $|\Psi| = \eta$, spontaneously breaks the remaining $U(1)$ symmetry, the symmetry can be restored at high temperatures where $\Psi$ is stabilized at zero.
In this case, Q-balls of $\Phi$ can form and remain stable, as in single-field models.

On the other hand, once $\Psi$ develops a nonzero VEV, the remaining $U(1)$ symmetry is spontaneously broken, and it becomes nontrivial whether the Q-ball can remain stable.
Suppose that $\Phi$ still constitutes a Q-ball even after the SSB. 
Inside the Q-ball, the direction $\propto -n_2 \theta_\Phi + n_1 \theta_\Psi$ remains massless, corresponding to the NG mode, and the orthogonal direction becomes massive due to the $\kappa$ term.
Here, $\theta_\Phi$ and $\theta_\Psi$ denote the complex phases of $\Phi$ and $\Psi$, respectively.
Although this massless NG mode exists, it cannot undergo stationary rotation inside the Q-ball for the following reason.
Let us first assume that the Q-ball rotates only in the massless direction, keeping the massive combination static.
This implies a time evolution of $\theta_\Psi$.
However, outside the Q-ball, $\Psi$ has a nonzero VEV throughout space, and both its temporal and spatial derivatives must be suppressed as $r \to \infty$ to ensure finite charge and energy.
As a result, the rotation of $\theta_\Psi$ inside the Q-ball cannot be coherent with the exterior, and the phase motion would propagate outward and carry away the conserved charge from the Q-ball.
Thus, for the Q-ball to remain a stable and localized object, only $\theta_\Phi$ can rotate, as in ordinary Q-balls, even though it is no longer a massless degree of freedom due to the $\kappa$ term.
As a result, the trajectory of $\Phi$ in the complex plane becomes perturbed,%
\footnote{
The perturbed trajectory resembles that in the presence of the explicit $U(1)$ breaking~\cite{Kawasaki:2005xc,Kawasaki:2019ywz,Kawasaki:2025}.
}
and $\Psi$ also acquires time dependence inside the Q-ball.
If the effect of the interaction term is sufficiently small, these perturbations remain mild, and the Q-ball may remain approximately stable.
On the other hand, a sizable interaction destabilizes the Q-ball.
As we will see in the next section, the fate of such unstable Q-balls depends on quantitative parameters such as the magnitude and the powers of the interaction term.

\section{Numerical results}
\label{sec: numerical}

To investigate the fate of a Q-ball, we numerically solve the equations of motion for the complex scalar fields. 
To make the system numerically tractable, we consider spherically symmetric configurations, $\Phi = \Phi(t,r)$ and $\Psi = \Psi(t,r)$.
In our numerical simulations, we use dimensionless variables defined by
\begin{gather}
    \tilde{t}
    \equiv 
    mt
    \ , ~~
    \tilde{r}
    \equiv 
    mr
    \ , ~~
    \tilde{\Phi} 
    \equiv 
    \frac{\Phi}{\Phi_c}
    \ , ~~
    \tilde{\Psi} 
    \equiv 
    \frac{\Psi}{\Phi_c}
    \ .
\end{gather}
Using these variables, we define the dimensionless potential as
\begin{align}
    \tilde{V}(\tilde{\Phi}, \tilde{\Psi})
    \equiv
    \frac{V(\Phi, \Psi)}{m^2 \Phi_c^2}
    =
    \left[ 1 - K \log\left( |\tilde{\Phi}|^2 + \epsilon \right) \right] |\tilde{\Phi}|^2
    +
    \frac{\tilde{\lambda}}{4} \left( |\tilde{\Psi}|^2 - \tilde{\eta}^2 \right)^2
    +
    \left( \tilde{\kappa} \tilde{\Phi}^{n_1} \tilde{\Psi}^{n_2} + \mathrm{h.c.} \right)
    \ ,
\end{align}
with 
\begin{align}
    \tilde{\lambda}
    \equiv 
    \frac{\Phi_c^2}{m^2} \lambda
    \ , ~~
    \tilde{\kappa} 
    \equiv 
    \frac{\Phi_c^{n_1 + n_2 - 2}}{m^2} \kappa
    \ , ~~
    \tilde{\eta}
    \equiv 
    \frac{\eta}{\Phi_c}
    \ .
\end{align}
Here, we introduce a small constant $0 < \epsilon \ll 1$ to regularize the potential near $|\tilde{\Phi}| \to 0$ in the numerical simulations.
Then, the equations of motion become
\begin{equation}
\begin{aligned}
    \ddot{\tilde{\Phi}}
    - \left( 
        \frac{\partial^2}{\partial \tilde{r}^2} 
        + \frac{2}{\tilde{r}} \frac{\partial}{\partial \tilde{r}}
    \right) \tilde{\Phi}
    +
    \frac{\partial \tilde{V}(\tilde{\Phi},\tilde{\Psi})}{\partial \tilde{\Phi}^*}
    &= 
    0
    \ , 
    \\ 
    \ddot{\tilde{\Psi}}
    - \left( 
        \frac{\partial^2}{\partial \tilde{r}^2} 
        + \frac{2}{\tilde{r}} \frac{\partial}{\partial \tilde{r}}
    \right) \tilde{\Psi}
    +
    \frac{\partial \tilde{V}(\tilde{\Phi},\tilde{\Psi})}{\partial \tilde{\Psi}^*}
    &= 
    0
    \ ,
\end{aligned}
\end{equation}
where the dots denote the derivatives with respect to $\tilde{t}$.
In the following, we omit the tildes for notational simplicity.

We now aim to simulate the evolution of the Q-ball during and after the SSB.
In a realistic cosmological setting, the $\Phi$ field begins to oscillate and form Q-balls, after which the $\Psi$ field develops a VEV and spontaneously breaks the $U(1)$ symmetry. However, in this case, the dynamics of $\Psi$ becomes highly nontrivial, making it difficult to investigate how the spontaneous breaking of $U(1)$ symmetry affects the time evolution of Q-balls due to the interaction between Q-balls and the $\Psi$ field excitations. Therefore, we initialize the $\Psi$ field with a constant expectation value from the beginning and gradually turn on the interaction between $\Phi$ and $\Psi$, in order to focus on the effects of the $U(1)$ symmetry breaking on Q-balls.
To this end, we introduce a time-dependent coupling while keeping the potential for $\Psi$ fixed. 
Specifically, we replace $\kappa$ with a time-dependent coupling $\kappa_t$ given by
\begin{align}
    \kappa_t(t)
    =
    \frac{\kappa}{2} 
    \left( 1 + \tanh \frac{t - t_c}{\Delta t} \right)
    \ ,
\end{align}
which smoothly increases from $\kappa_t \simeq 0$ for $t \ll t_c$ to $\kappa_t \simeq \kappa$ for $t \gg t_c$.
In the following, we set $t_c = 150$ and $\Delta t = 10\pi$.
With this choice, the interaction is negligibly small at the initial time, $t = t_i = 0$.
Therefore, we adopt the Q-ball solution in a single-field model as the initial condition:
\begin{equation}
\begin{gathered}
    \Phi(t_i,r)
    =
    e^{- r^2/R^2}
    , \quad 
    \dot{\Phi}(t_i,r)
    =
    i \omega e^{- r^2/R^2} 
    \ ,
    \\
    \Psi(t_i,r) 
    =
    \eta
    , \quad 
    \dot{\Psi}(t_i,r) 
    =
    0
    \ .
\end{gathered}
    \label{eq: initial}
\end{equation}

As for the boundary condition, we use the Neumann boundary condition at the inner boundary, $r = 0$, and the absorbing boundary condition~\cite{Salmi:2012ta} at the outer boundary, $r = r_\mathrm{max}$:
\begin{equation}
\begin{gathered}
    \partial_r \Phi|_{r = 0}
    =
    \partial_r \Psi|_{r = 0}
    =
    0
    \ ,
    \\
    \left[ 
        \partial_{r} \dot{\Phi}
        + \frac{\dot{\Phi}}{r}
        + \frac{1}{2} \frac{\partial V}{\partial \Phi}
    \right]_{r = r_\mathrm{max}}
    =
    \left[
        \partial_{r} \dot{\Psi}
        + \frac{\dot{\Psi}}{r}
        + \frac{1}{2} \frac{\partial V}{\partial \Psi}
    \right]_{r = r_\mathrm{max}}
    = 0
    \ .
\end{gathered}
\end{equation}

To quantify the field evolution, we define the $U(1)$ charges associated with $\Phi$ and $\Psi$ as
\begin{align}
    Q_\Phi 
    &\equiv 
    4\pi \int_0^{r_\mathrm{int}} \mathrm{d} r \, r^2 q_\Phi
    \equiv 
    4\pi \int_0^{r_\mathrm{int}} \mathrm{d} r \, r^2 
    i \left( \dot{\Phi}^* \Phi - \Phi^* \dot{\Phi}\right)
    \ ,
    \\
    Q_\Psi 
    &\equiv 
    4\pi \int_0^{r_\mathrm{int}} \mathrm{d} r \, r^2 q_\Psi
    \equiv 
    4\pi \int_0^{r_\mathrm{int}} \mathrm{d} r \, r^2 
    i \left( \dot{\Psi}^* \Psi - \Psi^* \dot{\Psi}\right)
    \ ,
    \label{eq: Psi charge}
\end{align}
where the upper limit of the integration, $r= r_\mathrm{int}$, is chosen appropriately in each case to evaluate the contribution from the Q-ball configuration while eliminating that from outgoing waves.
With the $\kappa$ term, the combination of $Q_\mathrm{con} \equiv n_2 Q_\Phi - n_1 Q_\Psi$ becomes the conserved charge, corresponding to the unbroken $U(1)$ symmetry.
In addition, we evaluate the total energy, defined as
\begin{align}
    E 
    &\equiv 
    4\pi \int_0^{r_\mathrm{int}} \mathrm{d} r \, r^2 \rho
    \ ,
    \label{eq: energy}
\end{align}
where the energy density  $\rho$ is given by
\begin{align}
    \rho 
    \equiv 
      |\dot{\Phi}|^2 + |\partial_r \Phi|^2
    + |\dot{\Psi}|^2 + |\partial_r \Psi|^2
    + V(\Phi, \Psi)
    \ .
\end{align}

In the following, we show the numerical results classifying the cases as $(n_1, n_2) = (1,3)$, $(2,2)$, and $(3,1)$.
While we vary $n_1$, $n_2$, and $\eta$, we fix
\begin{gather}
    K = 0.05 
    \ , \quad 
    \lambda = 1
    \ , \quad 
    \kappa = 0.1
    \ .
    \label{eq: baseline parameters}
\end{gather}
In this case, the initial configuration contains the energy and charge of 
\begin{align}
    E_\mathrm{ini} 
    \simeq 
    1121
    \ , \quad 
    Q_{\Phi,\mathrm{ini}}
    \simeq 
    1045 
    \ , \quad 
    Q_{\Psi,\mathrm{ini}}
    =
    0
    \ .
\end{align}

When we adopt larger values of \( n_1 \), such as \( (n_1, n_2) = (4, 1) \), we find that it becomes difficult to observe the effects of the SSB unless we extend our setup by introducing a term that stabilizes the potential. 
This is because the potential becomes unbounded from below due to the $\kappa$ term.  
Once we fix \( \Psi = \eta \), the effective potential for \( \Phi \) becomes shallow and unstable for large \( |\Phi| \), particularly along the direction where \( \Phi^{n_1} \) is real and negative.  
This instability becomes more pronounced for larger values of \( n_1 \).  
As a result, our analysis would be restricted to very small values of \( \eta \), which makes it difficult to probe the effects of the $\kappa$ term for \( n_1 \geq 4 \).

\subsection{\texorpdfstring{$(n_1,n_2)=(1,3)$}{}}
\label{subsec: n1 = 1}

First, we consider the case with $(n_1,n_2)=(1,3)$.
In this case, the potential minimum deviates from $(|\Phi|, |\Psi|) = (0, \eta)$, and the potential value at the minimum, $V_\mathrm{min}$, becomes negative.
To evaluate the excitation energy due to the initial Q-ball and its subsequent evolution, we subtract $V_\mathrm{min}$ from $\rho$:
\begin{align}
    E 
    &\equiv 
    4\pi \int_0^{r_\mathrm{int}} \mathrm{d} r \, r^2
    (\rho - V_\mathrm{min})
    \ ,
\end{align}
where we adopt $r_\mathrm{int} = 3R \simeq 19$, which is large enough to include the contribution from the localized condensate.

\begin{figure}[t]
    \centering
    \begin{minipage}{0.47\textwidth}
        \centering
        \includegraphics[width=\textwidth]{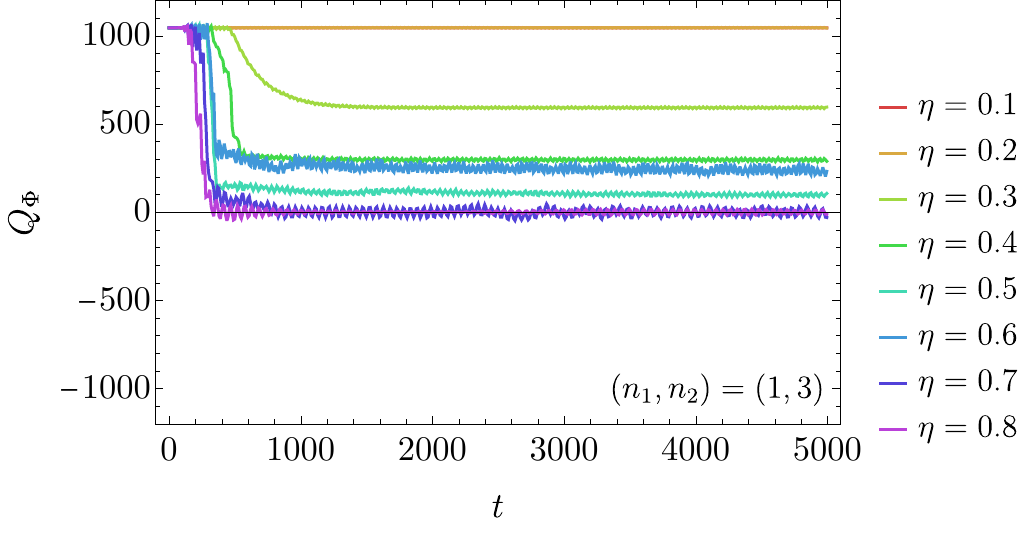}
    \end{minipage}
    \hspace{5mm}
    \begin{minipage}{0.47\textwidth}
        \centering
        \includegraphics[width=\textwidth]{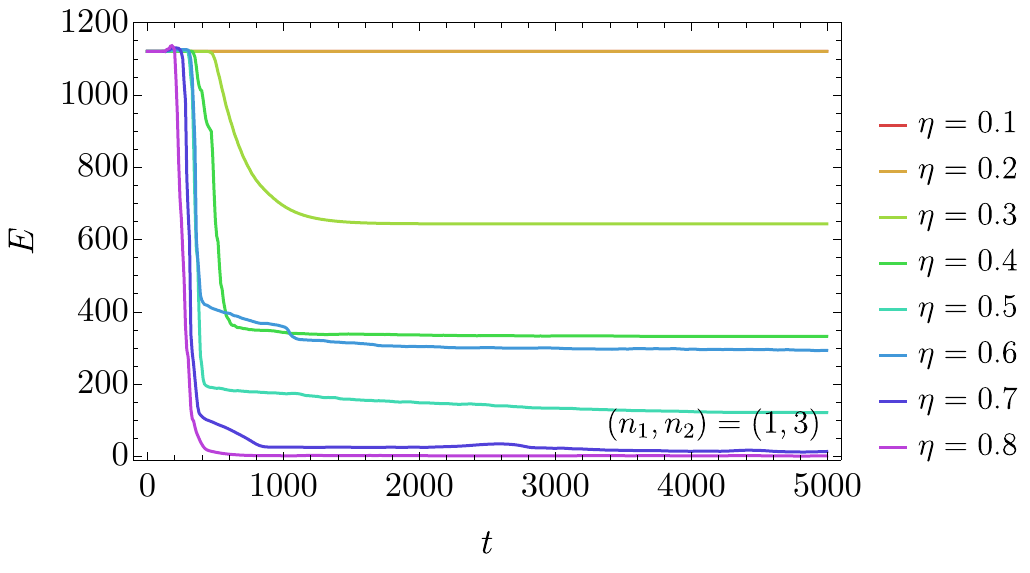}
    \end{minipage}
    \\
    \begin{minipage}{0.47\textwidth}
        \centering
        \includegraphics[width=\textwidth]{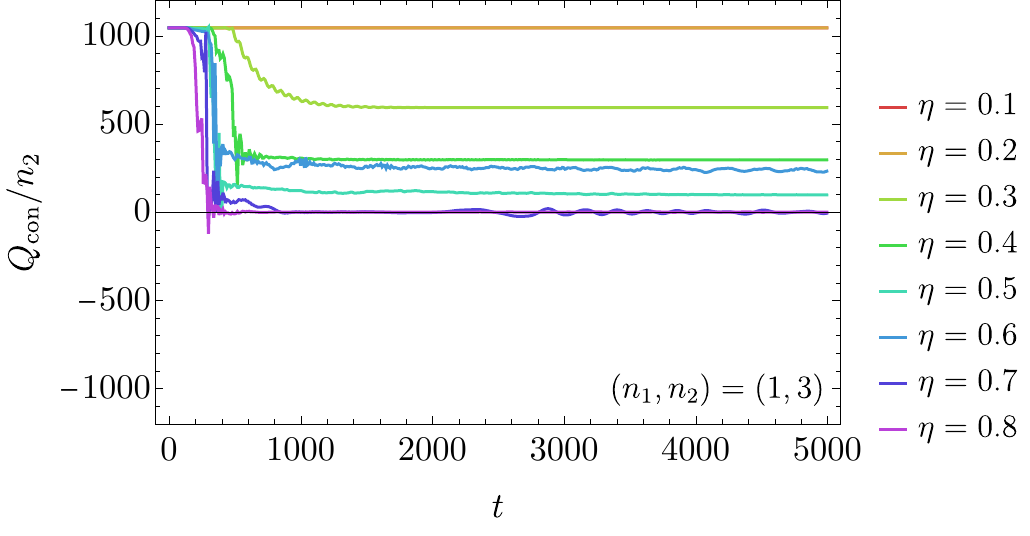}
    \end{minipage}
    \hspace{5mm}
    \begin{minipage}{0.47\textwidth}
        \centering
        \includegraphics[width=\textwidth]{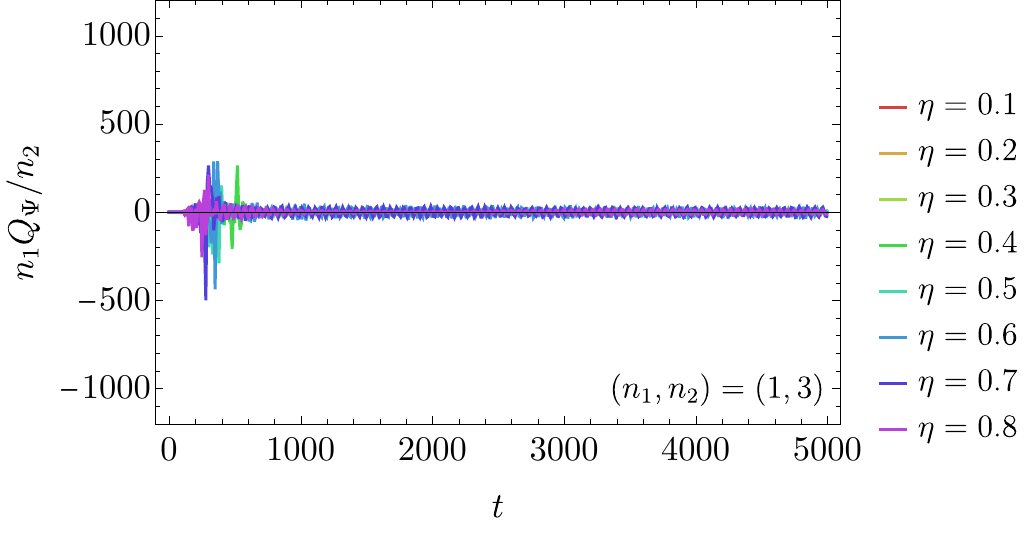}
    \end{minipage}
    \caption{%
        The time evolutions of $Q_\Phi$ (top left), $E$ (top right), $Q_\mathrm{con}/n_2$ (bottom left), and $n_1 Q_\Psi/n_2$ (bottom right) for $(n_1,n_2) = (1,3)$ and different values of $\eta$. 
        For large $\eta$, the Q-ball decays, releasing its charge, while for small $\eta$, it remains as a remnant with slightly reduced charge after emitting part of its charge.
    }
    \label{fig: n13 eta}
\end{figure}

In Fig.~\ref{fig: n13 eta}, we show the time evolution of $Q_\Phi$, $E$, $Q_\mathrm{con}$, and $Q_\Psi$ for different values of $\eta$. 
Here, $Q_\mathrm{con}$ and $Q_\Psi$ are normalized according to the relation  $Q_\Phi = Q_\mathrm{con}/n_2 + n_1 Q_\Psi/n_2$, and $Q_\mathrm{con}$ corresponds to the charge of the remaining $U(1)$ symmetry.
While both the charge and energy remain constant for $\eta \leq 0.2$, they decrease for larger $\eta$, and the Q-ball relaxes to a smaller, stable configuration.
For $\eta = 0.8$, the Q-ball completely decays, leaving no remnant.
These results indicate that a Q-ball can survive in the presence of the spontaneous breaking of $U(1)$ symmetry, as long as its effect is not too large.
Although $Q_\Phi$ rapidly oscillates for some $\eta$, the behavior of $Q_\mathrm{con}$ is more stable due to the underlying $U(1)$ symmetry. Note that the charge and energy at the final time do not exhibit a monotonic dependence on $\eta$.
In particular, the remaining charge and energy are larger for $\eta = 0.6$ compared to the case with $\eta = 0.5$.

\begin{figure}[t]
    \centering
    \begin{minipage}{0.45\textwidth}
        \centering
        \includegraphics[width=\textwidth]{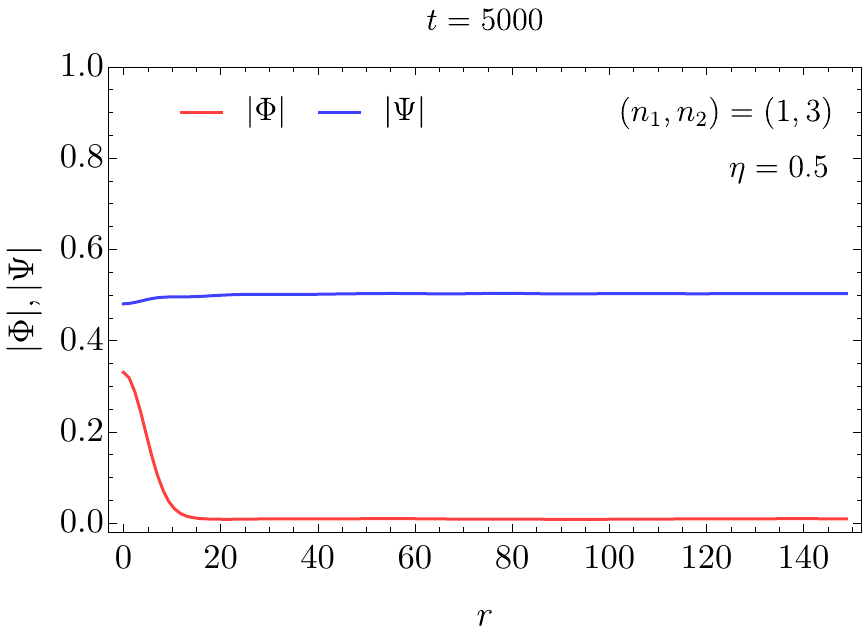}
    \end{minipage}
    \hspace{5mm}
    \begin{minipage}{0.45\textwidth}
        \centering
        \includegraphics[width=0.7\textwidth]{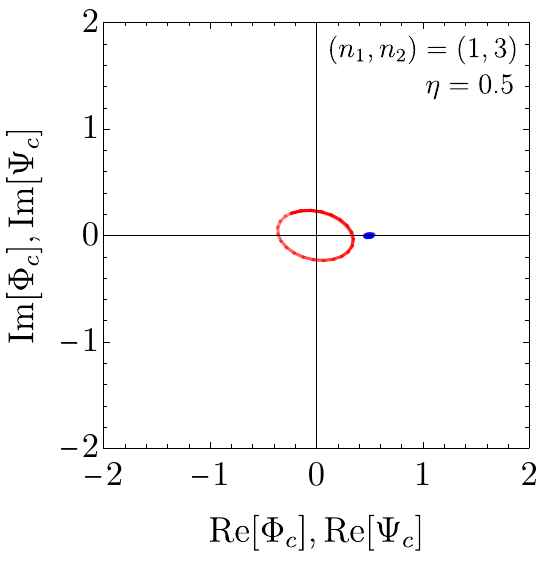}
    \end{minipage}
    \\
    \begin{minipage}{0.45\textwidth}
        \centering
        \includegraphics[width=\textwidth]{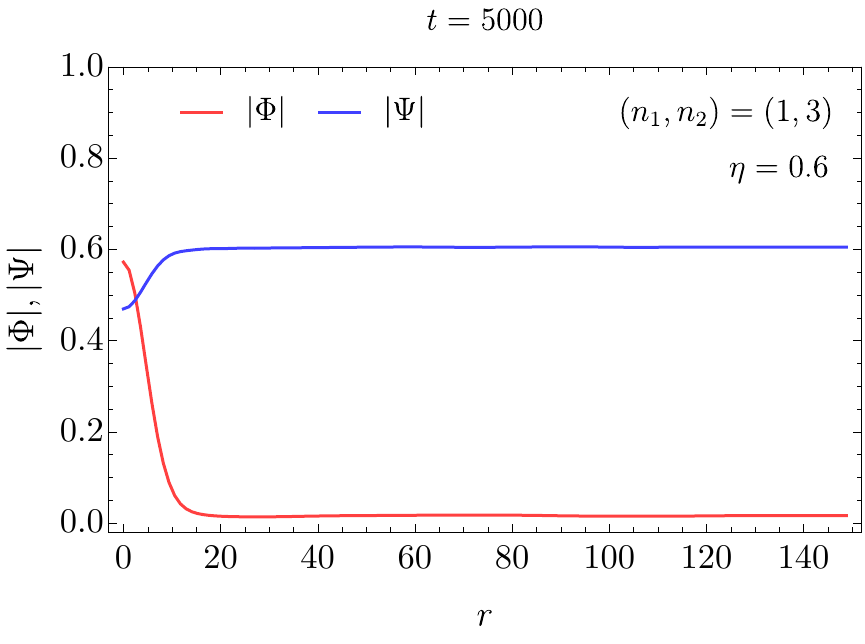}
    \end{minipage}
    \hspace{5mm}
    \begin{minipage}{0.45\textwidth}
        \centering
        \includegraphics[width=0.7\textwidth]{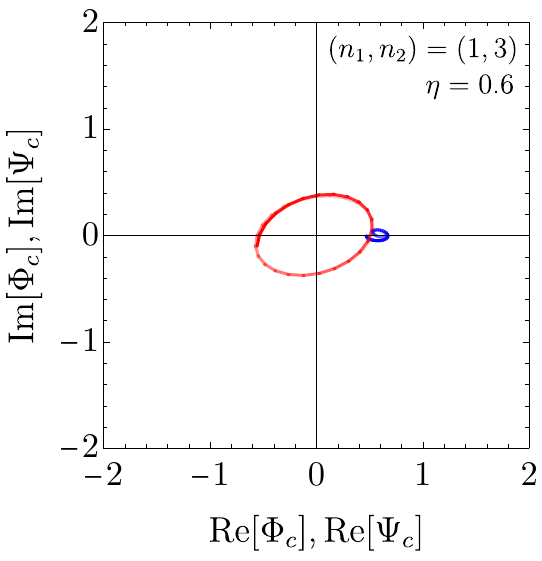}
    \end{minipage}
    \caption{%
        The field configurations and trajectories well after the $\kappa$ term is turned on for $(n_1,n_2) = (1,3)$.
        We adopt $\eta = 0.5$ for the top row and $\eta = 0.6$ for the bottom row.
        \textit{Left panels}: the spatial configuration of $|\Phi|$ and $|\Psi|$ at $t = 5000$.
        \textit{Right panels}: the trajectory of $\Phi$ and $\Psi$ at the origin for $4990 \leq t \leq 5000$.
        The trajectories are shown in lighter colors for earlier times.
    }
    \label{fig: n13 final}
\end{figure}

To clarify what remains at late times, we show the field configurations at $t = 5000$ and the field trajectories at $r = 0$ for $t \sim 5000$ in Fig.~\ref{fig: n13 final}.
Here, we consider cases with $\eta = 0.5$ and $0.6$.
While the spatial profiles of $|\Phi|$ exhibit similar shapes for both values of $\eta$, $|\Psi|$ is more perturbed for $\eta = 0.6$.
This tendency is also evident in the field trajectory.
From this difference, we expect that the significantly large $Q_\Phi$ observed at $\eta = 0.6$ in Fig.~\ref{fig: n13 eta} is due to the interplay of $\Phi$ and $\Psi$ fields such as synchronized oscillations.
We also consider a situation where $|\Psi|$ is fixed to $\eta$ due to large $\lambda$.
We find that the energy at the final time monotonically depends on $\eta$, supporting the above interpretation.
See Fig.~\ref{fig: n13H eta} in Appendix~\ref{app: heavy limit} for details.

\subsection{\texorpdfstring{$(n_1,n_2)=(2,2)$}{}}
\label{subsec: n1 = 2}

\begin{figure}[htbp]
    \centering
    \begin{minipage}{0.47\textwidth}
        \centering
        \includegraphics[width=\textwidth]{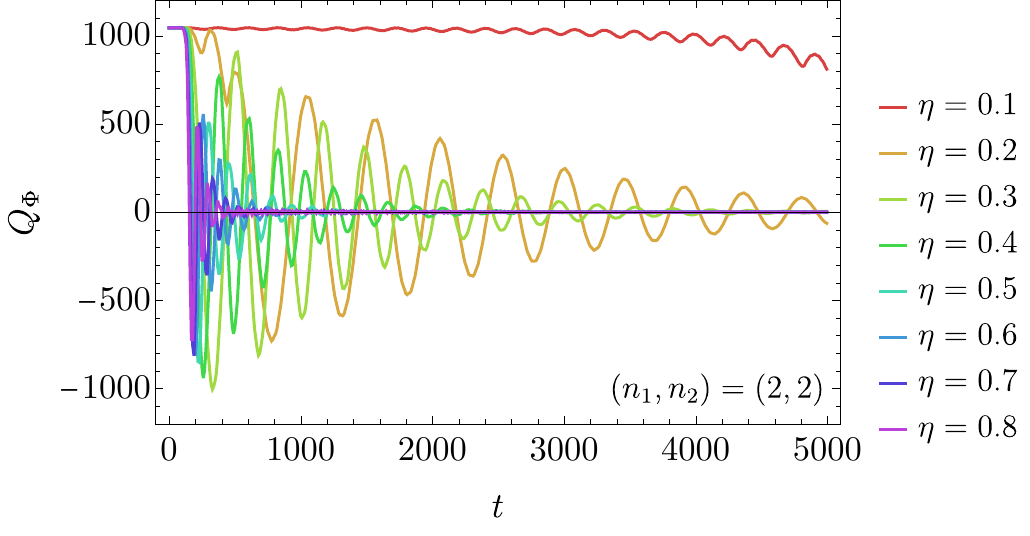}
    \end{minipage}
    \hspace{5mm}
    \begin{minipage}{0.47\textwidth}
        \centering
        \includegraphics[width=\textwidth]{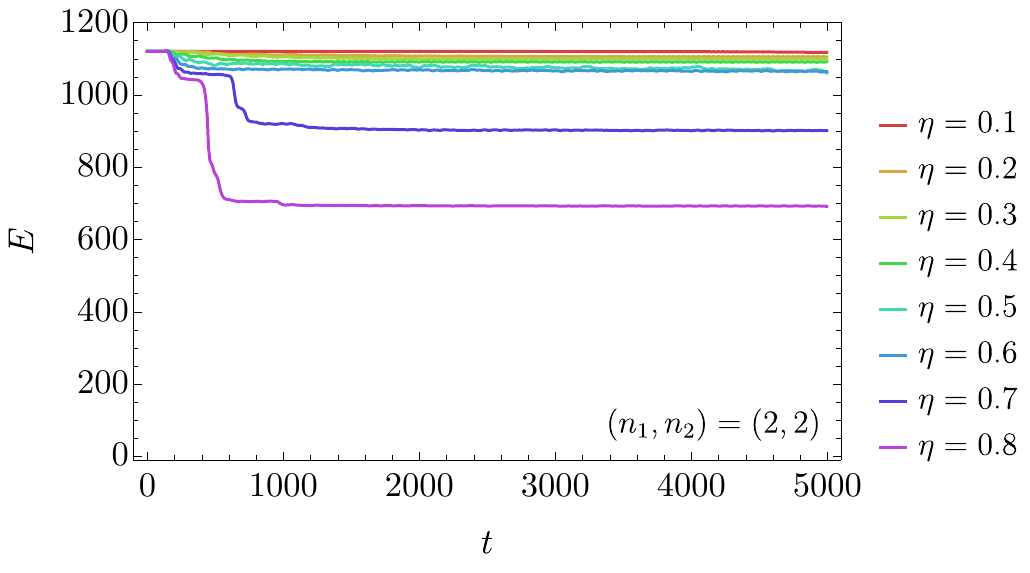}
    \end{minipage}
    \\
    \begin{minipage}{0.47\textwidth}
        \centering
        \includegraphics[width=\textwidth]{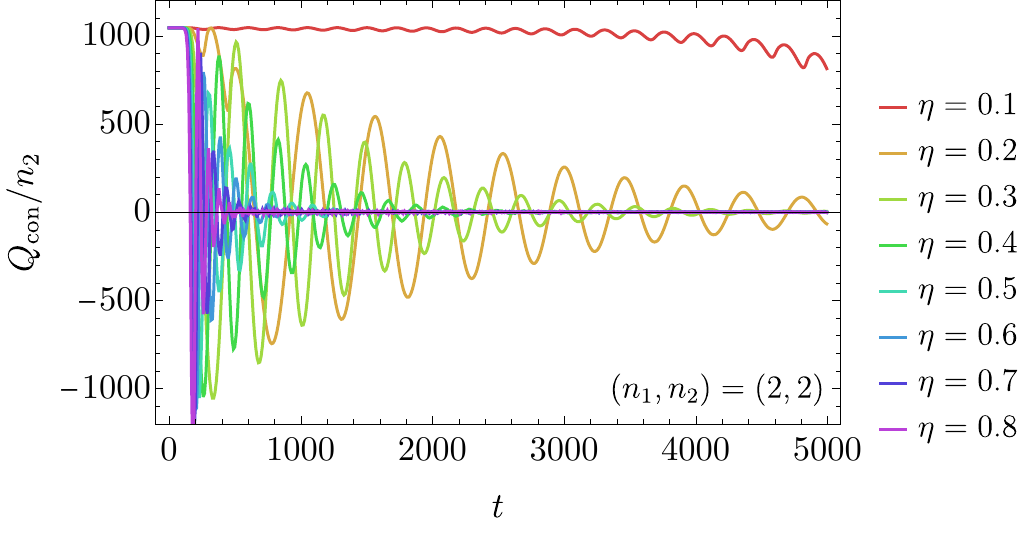}
    \end{minipage}
    \hspace{5mm}
    \begin{minipage}{0.47\textwidth}
        \centering
        \includegraphics[width=\textwidth]{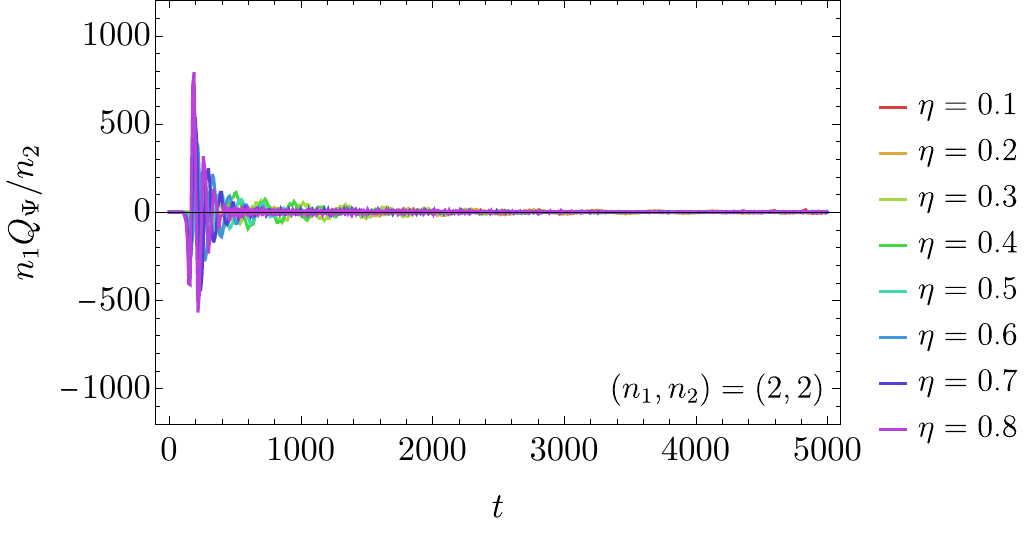}
    \end{minipage}
    \caption{%
        Same as Fig.~\ref{fig: n13 eta} except for $(n_1, n_2) = (2,2)$. 
        Although the charge efficiently escapes from the Q-ball, the energy remains confined in the form of oscillons/I-balls.
    }
    \label{fig: n22 eta}
\end{figure}

Next, we consider the case with $(n_1,n_2)=(2,2)$.
In Fig.~\ref{fig: n22 eta}, we show the time evolution of the charges and energy with $r_\mathrm{int} = 3R$.
We can see that $Q_\Phi$ and $Q_\mathrm{con}$ decrease for any $\eta$ and completely vanish within the simulation time for large $\eta \,(\ge 0.3)$.
On the other hand, the energy density remains nonzero and becomes almost constant at late times.
These results indicate that the Q-ball effectively releases the charge and transforms into a compact object with a negligibly small charge.

\begin{figure}[htbp]
    \centering
    \begin{minipage}{0.45\textwidth}
        \centering
        \includegraphics[width=\textwidth]{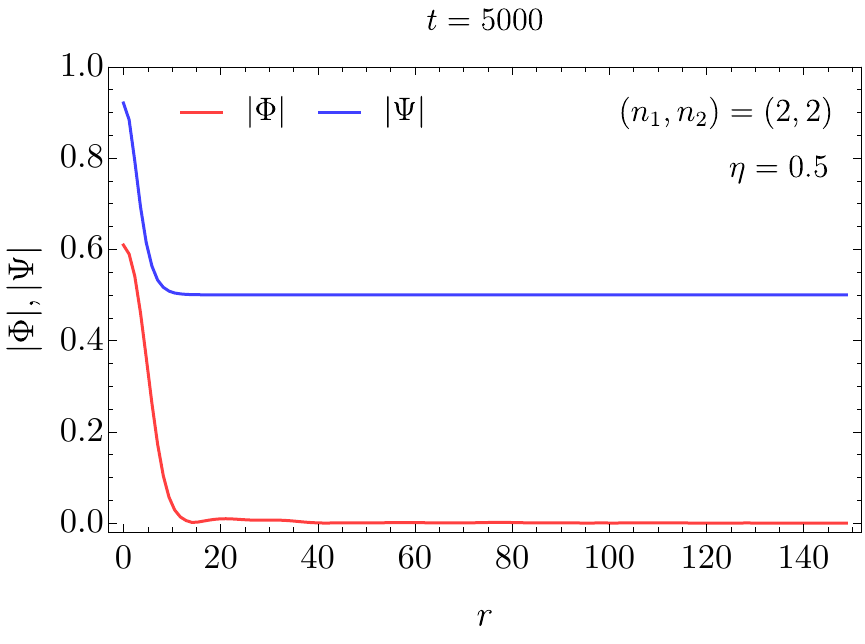}
    \end{minipage}
    \hspace{5mm}
    \begin{minipage}{0.45\textwidth}
        \centering
        \includegraphics[width=0.7\textwidth]{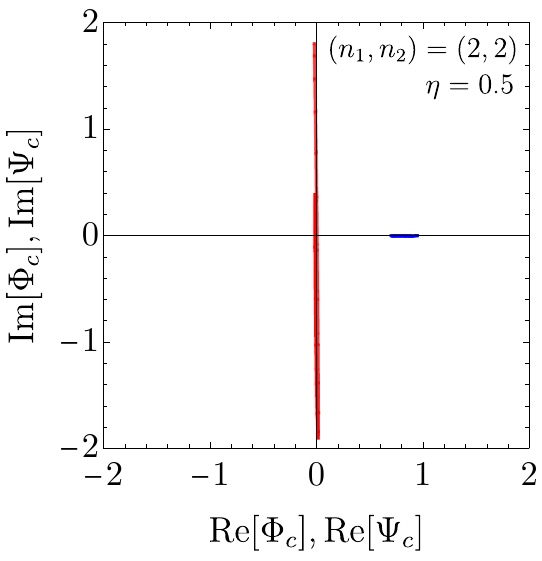}
    \end{minipage}
    \caption{%
        Same as Fig.~\ref{fig: n13 final} except for $(n_1, n_2) = (2,2)$ and $\eta = 0.5$.
        After the charge has leaked out, $\Phi(\Psi)$ oscillates along a single direction inside the oscillon.
    }
    \label{fig: n22 final}
\end{figure}

We show the field configuration and the field trajectories at $r = 0$ for $\eta = 0.5$ in Fig.~\ref{fig: n22 final}.
Both $\Phi$ and $\Psi$ are excited near the center, and their trajectories are aligned with the imaginary and real axes, respectively.
When focusing on the behavior of $\Phi$,  this structure can be interpreted as an oscillon/I-ball~\cite{Bogolyubsky:1976yu,Gleiser:1993pt,Kasuya:2002zs}, which often arises in real scalar field theories.
The appearance of an oscillon can be understood as follows.
If we assume $\Psi$ to be fixed at $\eta$, the $\kappa$ term results in a trough along the imaginary axis of $\Phi$ for $\kappa > 0$.
Due to the existence of this trough, $\Phi$ comes to oscillate along the imaginary axis after $\kappa$ is turned on.
Since the potential for $\Phi$ with a logarithmic correction admits an oscillon solution~\cite{Kasuya:2002zs}, the Q-ball finally becomes an oscillon.
Since $\Psi$ is dynamical, it receives backreaction from $\Phi$ through the $\kappa$ term and oscillates in a similar manner.

Indeed, the time evolution of the energy exhibits a step-like behavior for $\eta = 0.7$ and $0.8$, which is a characteristic feature of oscillons.
Unlike Q-balls, oscillons do not carry a conserved charge associated with a global symmetry. Their longevity instead relies on the approximate conservation of the adiabatic invariant~\cite{Kasuya:2002zs,Kawasaki:2015vga}.
Then, it gradually decays radiating high-frequency waves of the scalar field.
Its decay rate depends on the powers of the radiation modes, which oscillate depending on the oscillon mass.
When the power of the primary radiating mode vanishes, an oscillon decays only via higher-frequency modes, resulting in a long-lived state.
Once the oscillon mass decreases enough and the decay by the primary mode becomes efficient again, the oscillon rapidly decays to the next long-lived state.
See, e.g., Refs.~\cite{Mukaida:2016hwd,Ibe:2019vyo,Ibe:2019lzv,Zhang:2020bec} for more detailed discussions.

\subsection{\texorpdfstring{$(n_1,n_2)=(3,1)$}{}}
\label{subsec: n1 = 3}

\begin{figure}[t]
    \centering
    \begin{minipage}{0.47\textwidth}
        \centering
        \includegraphics[width=\textwidth]{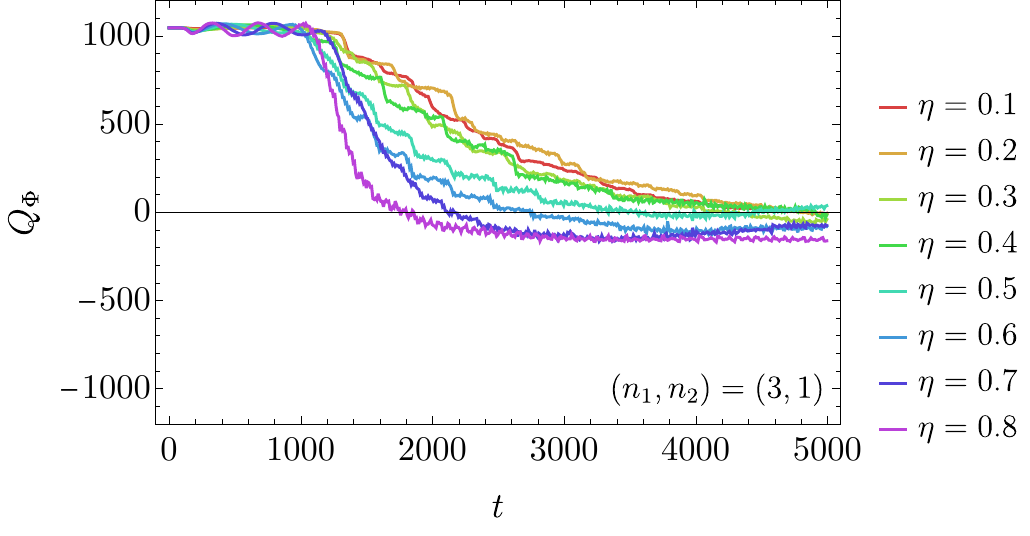}
    \end{minipage}
    \hspace{5mm}
    \begin{minipage}{0.47\textwidth}
        \centering
        \includegraphics[width=\textwidth]{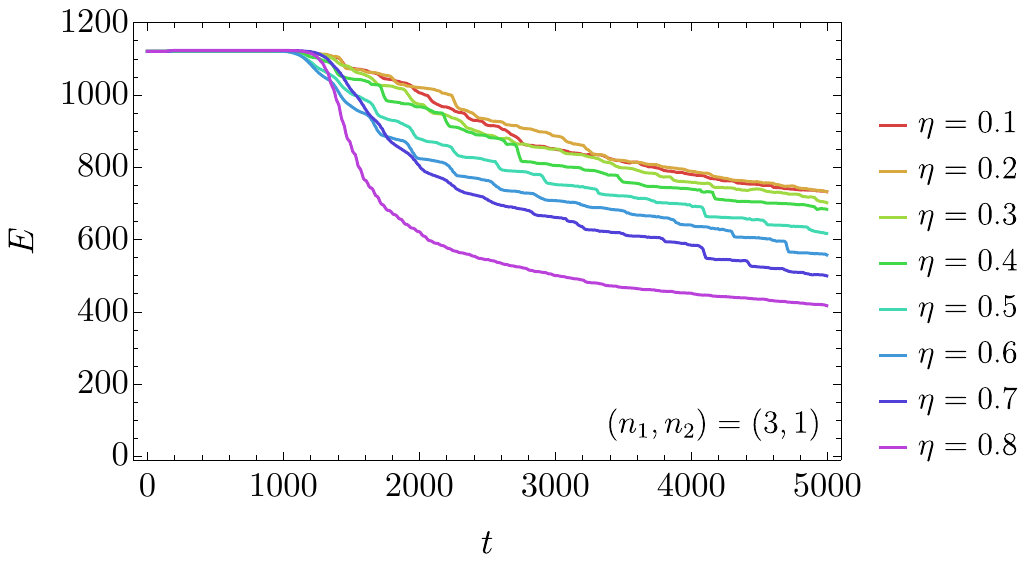}
    \end{minipage}
    \\
    \begin{minipage}{0.47\textwidth}
        \centering
        \includegraphics[width=\textwidth]{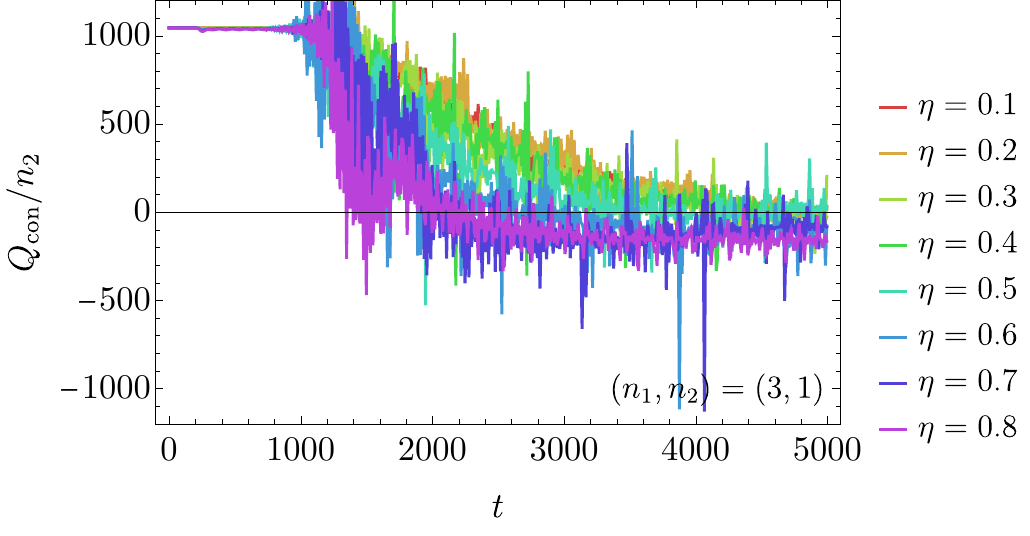}
    \end{minipage}
    \hspace{5mm}
    \begin{minipage}{0.47\textwidth}
        \centering
        \includegraphics[width=\textwidth]{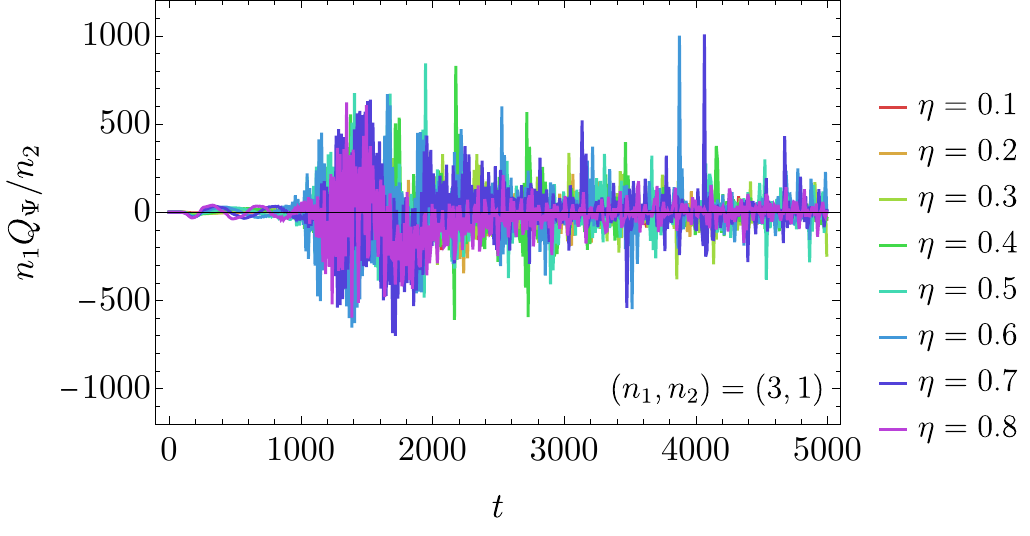}
    \end{minipage}
    \caption{%
        Same as Fig.~\ref{fig: n13 eta} except for $(n_1, n_2) = (3,1)$. 
        Both the charge and energy decrease significantly, but for large $\eta$, the charge appears to asymptote to a negative value.
    }
    \label{fig: n31 eta}
\end{figure}

\begin{figure}[t]
    \centering
    \begin{minipage}{0.45\textwidth}
        \centering
        \includegraphics[width=\textwidth]{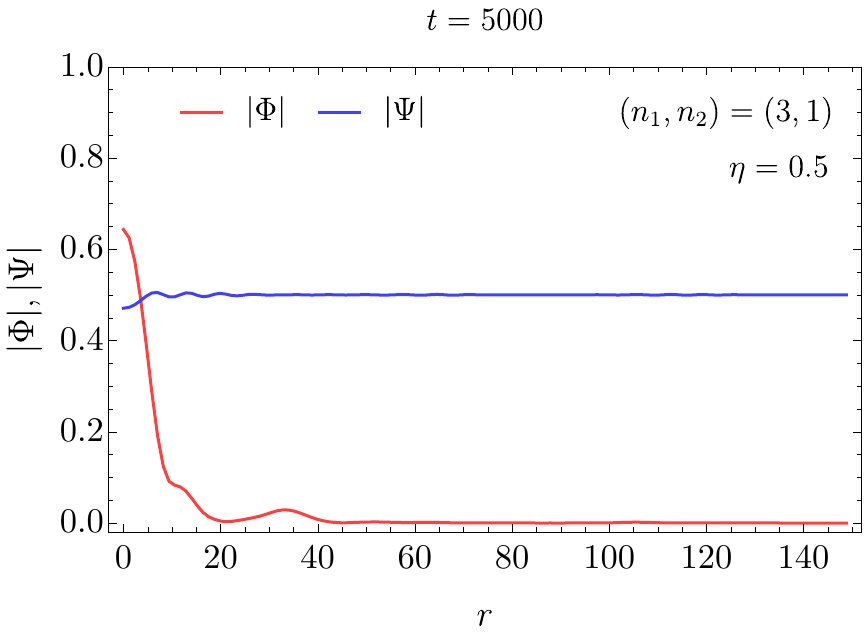}
    \end{minipage}
    \hspace{5mm}
    \begin{minipage}{0.45\textwidth}
        \centering
        \includegraphics[width=0.7\textwidth]{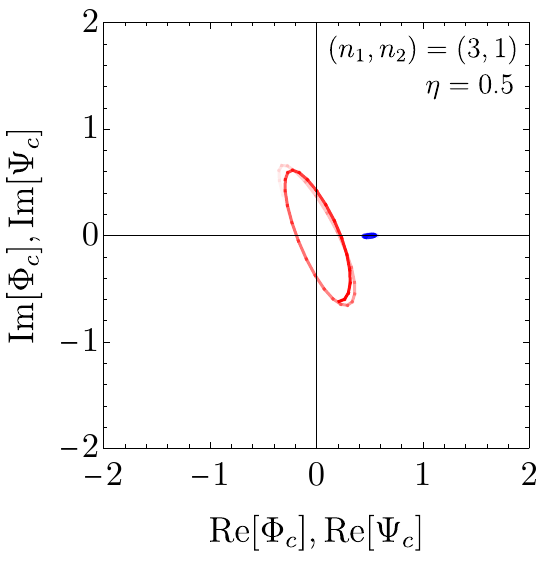}
    \end{minipage}
    \caption{%
        Same as Fig.~\ref{fig: n13 final} except for $(n_1, n_2) = (3,1)$.
    }
    \label{fig: n31 final}
\end{figure}

Finally, we consider the case with $(n_1,n_2)=(3,1)$.
We show the time evolution of the charges and energy in Fig.~\ref{fig: n31 eta}.
Here, we adopt $r_\mathrm{int} = 15 R \simeq 95$.
If we chose much smaller $r_\mathrm{int}$, $Q_\Phi$ and $E$ would not monotonically decrease since the structure of $\Phi$ spreads out of the integration range (see the left panel of Fig.~\ref{fig: n31 final}).
For all $\eta$ shown there, $Q_\Phi$ is perturbed for $t \gtrsim t_c$ and decays after some oscillations.
Note that $Q_\mathrm{con}$ remains almost constant for $t \lesssim 1000$ while $Q_\Phi$ oscillates.
Although $Q_\Phi$ and $E$ start to decrease at similar times for any $\eta$, they decrease more rapidly for larger $\eta$.
Note that $Q_\Psi$ significantly oscillates once the Q-ball starts to decay.
This comes from the outgoing waves of the phase component of $\Psi$.
Although free-propagating waves decrease its amplitude as $\propto r^{-1}$, $\Psi$ itself is almost constant outside the Q-ball.
As a result, the outgoing waves contribute to $q_\Psi$ as $r^{-1}$ (see Eq.~\eqref{eq: Psi charge}), and the integrated charge $Q_\Psi$ grows as $\propto r_\mathrm{int}$.
Since we now take a relatively large value of $r_\mathrm{int}$, the oscillations in $Q_\Psi$ and $Q_\mathrm{con}$ become more evident.

We show the field configuration and the field trajectories at $r = 0$ for $\eta = 0.5$ in Fig.~\ref{fig: n31 final}.
While $\Phi$ concentrates near the center, its shape is perturbed compared with the initial Gaussian configuration, and the trajectory significantly deviates from a circle.
$\Psi$ is also perturbed near the center with smaller amplitudes.
As shown in Fig.~\ref{fig: n31 eta}, $Q_\Phi$ becomes much smaller than its initial value at $t = 5000$, while $E$ remains about half of its initial value.
This suppression of $Q_\Phi$ results from cancellation between different radial regions.
In Fig.~\ref{fig: n31 final}, for example, $\Phi$ has a negative charge density for $r \lesssim 20$ and a positive charge density for $20 \lesssim r \lesssim 40$.%
\footnote{Such a counter-rotating region is also observed in a different setup with a spontaneously broken $U(1)$ symmetry~\cite{Fedderke:2025sic}.}

\begin{figure}[t]
    \centering
    \begin{minipage}{0.47\textwidth}
        \centering
        \includegraphics[width=\textwidth]{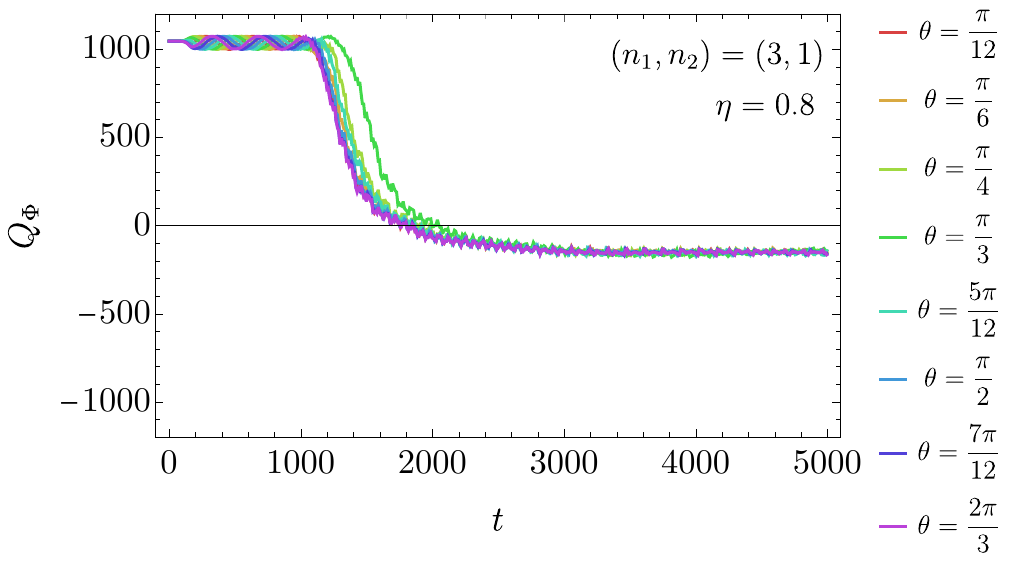}
    \end{minipage}
    \hspace{5mm}
    \begin{minipage}{0.47\textwidth}
        \centering
        \includegraphics[width=\textwidth]{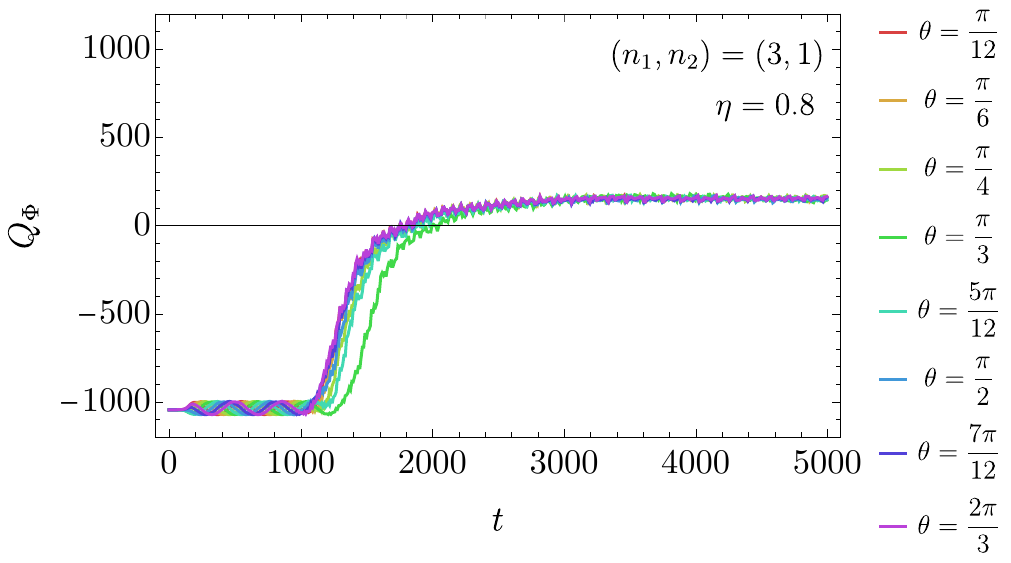}
    \end{minipage}
    \caption{%
        The time evolutions of $Q_\Phi$ for $(n_1,n_2) = (3,1)$ and different values of $\theta$ with the positive (left) and negative (right) initial charges of the Q-ball.
    }
    \label{fig: n31 charge initial}
\end{figure}

Note that $Q_\Phi$ finally approaches a certain negative value for, e.g., $\eta = 0.8$ although $Q_{\Phi,\mathrm{ini}}$ is positive.
To see if this relies on the specific initial condition, we also simulate the field dynamics for different initial conditions changing the initial phase and the charge sign.
Specifically, we set the initial conditions for $\Phi$ as
\begin{align}
    \Phi(t_i,r)
    =
    e^{- r^2/R^2 + i \theta}
    , \quad 
    \dot{\Phi}(t_i,r)
    =
    \pm i \omega e^{- r^2/R^2 + i \theta} 
    \ ,
\end{align}
while using the same initial conditions as Eq.~\eqref{eq: initial} for $\Psi$.
The plus and minus signs of $\dot{\Phi}$ correspond to the positive and negative initial charge of the Q-ball, respectively.
Considering $(n_1, n_2) = (3,1)$, we can identify $\theta \sim \theta + 2\pi/3$ by the redefinition of the complex phase of $\Phi$.
We show the time evolution of the charge for different values of $\theta$ with the positive and negative charge of the Q-ball in Fig.~\ref{fig: n31 charge initial}.
We find that, while the intermediate evolution depends on the initial conditions, the charge finally converges to a certain value.
This implies that there is an attractor configuration with a negative charge.

\section{Summary}

In this paper, we have investigated the evolution of Q-balls in the presence of spontaneous breaking of a global $U(1)$ symmetry. The stability of Q-balls relies on the conservation of global $U(1)$ charge, but when the symmetry is spontaneously broken, a massless NG boson emerges and provides a channel through which the charge can escape. This process may significantly affect the long-term behavior of Q-balls and their role in cosmology.
To examine the effects of this symmetry breaking, we studied a system of two complex scalar fields, $\Phi$ and $\Psi$: the $\Phi$ field is responsible for forming the Q-ball and the $\Psi$ field develops a VEV that spontaneously breaks the $U(1)$ symmetry. This setup allows us to isolate the impact of SSB on Q-ball evolution in a simple and controlled manner.

Specifically, we considered the interaction of $\Phi^{n_1} \Psi^{n_2} + \mathrm{h.c.}$ with $(n_1, n_2) = (1,3)$, $(2,2)$, and $(3,1)$.
In all cases, we observed that the Q-ball loses its charge and energy more significantly for larger values of the VEV of $\Psi$. On the other hand, for relatively small $\eta$, stable and localized objects often survived the evolution. 
The final configuration after the evolution, however,  is qualitatively different for different choices of $(n_1,n_2)$.

In the case of $(n_1,n_2) = (1,3)$, the Q-ball relaxes into a smaller Q-ball.
The charge of the remaining Q-ball does not exhibit a monotonic dependence on $\eta$, which may come from the synchronized oscillation of $\Phi$ and $\Psi$.
In the case of $(n_1,n_2) = (2,2)$, the Q-ball is deformed into an oscillon/I-ball.
In the case of $(n_1,n_2) = (3,1)$, the Q-ball decays at a certain time after the interaction term is turned on.
While the charge decreases rapidly, the energy decreases more gradually.
Interestingly, the final charge approaches a specific value regardless of the initial phase of the Q-ball.
This may suggest the existence of an attractor solution.

In these results, both \( \Phi \) and \( \Psi \) are dynamically perturbed due to their mutual interaction, i.e., the backreaction between the two fields modifies their trajectories and field values compared to the cases where \( \Phi \) forms a Q-ball in isolation or \( \Psi \) acquires a constant VEV independently.  
To clarify the importance of the dynamical degrees of freedom of \( \Psi \), we also consider a setup in which the radial component of \( \Psi \) is effectively frozen by fixing \( |\Psi| = \eta \), so that only the phase degree of freedom remains.
As a result, we find that the evolution of Q-balls is altered in the absence of the radial dynamics of \( \Psi \), although certain qualitative features remain similar for the same choice of \( (n_1, n_2) \).  
In particular, the charge dissipation still occurs, indicating that the presence of a massless NG mode outside the Q-ball—arising from the spontaneous breaking of the global \( U(1) \) symmetry—is the key factor governing the Q-ball decay. 
The detailed results of this analysis are presented in Appendix~\ref{app: heavy limit}.

The purpose of this work is to show the variety of the evolution of a Q-ball in a system with the spontaneously broken $U(1)$ and to establish a basis for further research on their cosmological implications.
However, we briefly mention possible cosmological implications.
In our analysis, we have assumed Q-balls as initial conditions without any explicit or spontaneous breaking of the associated global $U(1)$ symmetry.
Nevertheless, our results can be applied to Q-balls formed via the AD mechanism where the $U(1)$ symmetry is explicitly broken by the so-called A-terms.
This is because A-terms typically originate from higher-dimensional operators suppressed by a high-energy scale, and their effects become negligibly small once the field amplitude decreases in the expanding universe. By the time Q-balls form, the effect of the A-term has typically become negligible, as it affects the dynamics only near the onset of oscillations of the corresponding flat direction.
Furthermore, even in the absence of $U(1)$ charge asymmetry, the oscillating scalar field is known to evolve into Q-balls with both positive and negative charges~\cite{Kasuya:1999wv,Enqvist:2002si}.
One can also consider a scenario involving multiple flat directions, where Q-balls are formed from the oscillation of one flat direction, and at a later stage, another flat direction develops a nonzero VEV. This leads to spontaneous $U(1)$ symmetry breaking, which can give rise to the types of effects discussed in this work.
Thus, we can discuss the cosmological implications of our findings for a broad range of Q-ball formation mechanisms.
For instance, when Q-balls decay or deform to smaller Q-balls or oscillons, their charge and energy are radiated to the universe.
While these radiations can modify the thermal history of the universe or affect cosmological observables such as big bang nucleosynthesis, cosmic microwave background, and large-scale structure, the remaining objects can serve as dark matter.
In addition, gravitational waves can be emitted at that time although it requires deviation from spherical symmetry in the evolution of Q-balls.
Further investigation in these directions is left for future studies.

Finally, we comment on possible extensions of our analysis.
First, our analysis assumes spherical symmetry of the system and adopts the ideal Q-ball configuration as the initial condition.
To study more realistic evolution including the Q-ball formation, we need to conduct three-dimensional lattice simulations.
Second, we studied the renormalizable interaction terms and found the various results for different powers.
We can also study higher-order operators, including the $U(1)$ conserving and breaking ones, which are well-studied in the context of supersymmetric models.

Third, in our analysis, the spontaneous breaking of the remaining $U(1)$ symmetry by $\Psi$ influences the dynamics of $\Phi$ through the interaction term (the $\kappa$ term), modifying the potential structure. One can also consider more drastic effects where the VEV of $\Psi$ significantly alters the potential of $\Phi$. For instance, if the potential for $\Phi$ becomes steeper than a quadratic potential for all phases of $\Phi$, neither Q-balls nor oscillons are expected to survive. Also, there may be cases where the VEV of $\Psi$ induces a nonzero VEV for $\Phi$. In fact, we have observed such behavior in the case of $(n_1,n_2) = (1,3)$. A more extreme example is an interaction like $|\Phi|^2|\Psi|^2$, which can destabilize the origin of $\Phi$. 
Another possible extension of the present work is to explore different forms of the potential for $\Phi$. While we have adopted a mass term with a negative logarithmic correction, Q-ball solutions are known to exist for a broader class of potentials that are flatter than the quadratic one. In such cases, it is nontrivial whether oscillons remain stable after the spontaneous breaking of the U(1) symmetry, since their longevity is due to the adiabatic invariant which is specific to the quadratic potential~\cite{Kasuya:2002zs}.
Understanding how robust Q-balls are in such cases is crucial for studying their cosmological evolution. 
We leave these issues for future work.

\section*{Note Added}
During the early stages of this work, we became aware that another group was independently investigating a related subject~\cite{Kobayashi:2025qao}. After discussing with them, we agreed to coordinate and submit our papers on the same day. We appreciate the helpful interaction.
While our study focuses on a two-field setup, Ref.~\cite{Kobayashi:2025qao} considers a single-field model with spontaneous U(1) symmetry breaking, in a different context.

\section*{Acknowledgments}
We thank Jin Kobayashi, Kazunori Nakayama, and Masaki Yamada for valuable discussions. 
We also acknowledge the workshop ``What is dark matter? - Comprehensive study of the huge discovery space in dark matter", held at the Yukawa Institute for Theoretical Physics, where the present work was initiated.
This work is supported by JSPS Core-to-Core Program (grant number: JPJSCCA20200002) (F.T.), JSPS KAKENHI Grant Numbers 20H01894 (F.T.), 20H05851 (F.T., M.K.), 21K03567 (M.K.), 23KJ0088 (K.M.),  24K17039 (K.M.),  25H02165 (F.T.),  and 25K07297 (M.K.).
This article is based upon work from COST Action COSMIC WISPers CA21106, supported by COST (European Cooperation in Science and Technology).

\appendix

\section{Heavy limit of \texorpdfstring{$|\Psi|$}{}}
\label{app: heavy limit}

To clarify the effect of the degree of freedom of $\Psi$, we investigate the case where the radial component of $\Psi$ is effectively fixed to $|\Psi| = \eta$ by taking extremely large $\lambda = 10^4$.

\begin{figure}[t]
    \centering
    \begin{minipage}{0.47\textwidth}
        \centering
        \includegraphics[width=\textwidth]{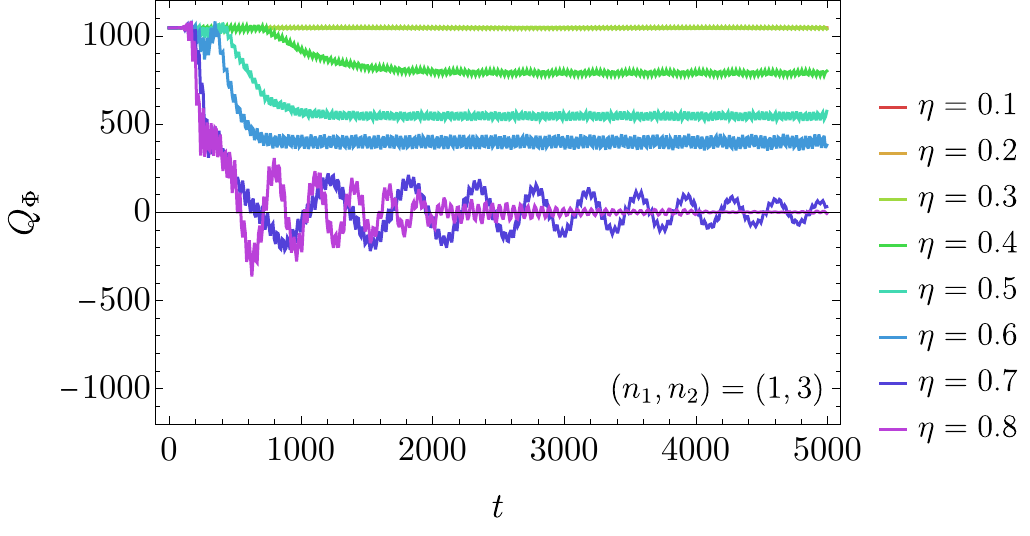}
    \end{minipage}
    \hspace{5mm}
    \begin{minipage}{0.47\textwidth}
        \centering
        \includegraphics[width=\textwidth]{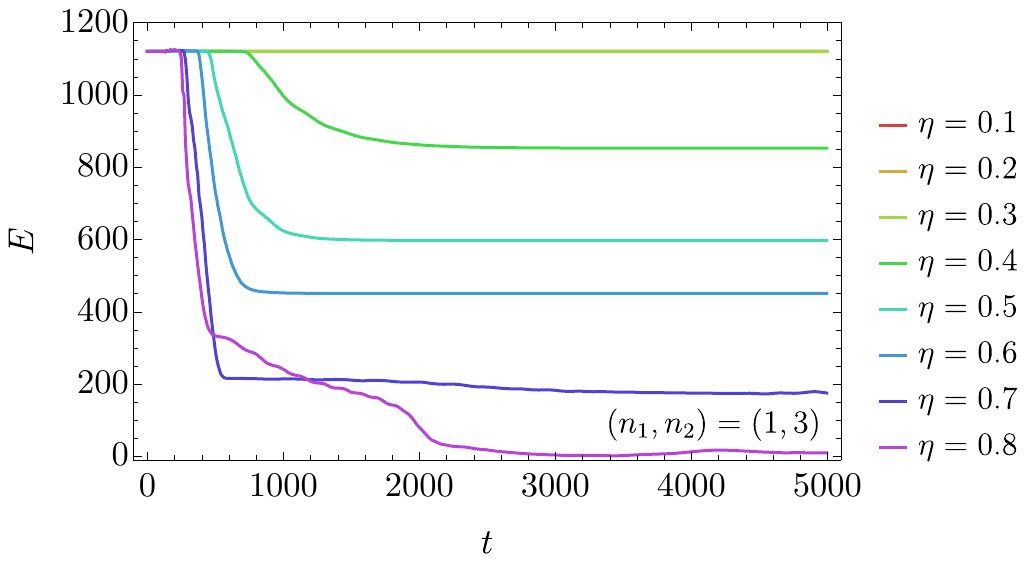}
    \end{minipage}
    \caption{%
        Same as the upper panels in Fig.~\ref{fig: n13 eta} except for fixing $|\Psi| = \eta$.
    }
    \label{fig: n13H eta}
\end{figure}

We show the results for $(n_1, n_2) = (1,3)$ in Fig.~\ref{fig: n13H eta}.
Compared with the results in the main text, the initial Q-ball is stable for larger $\eta$, and $Q_\Phi$ shows oscillations for large $\eta$. 
While $Q_\Phi$ exhibits small and rapid oscillations on large and slow ones, $Q_\mathrm{con}$ exhibits only the slow oscillations.
In contrast to the results in the main text, the final energy and charge monotonically depend on $\eta$, which implies that the non-monotonic dependence shown in Fig.~\ref{fig: n13 eta} relies on the radial component of $\Psi$. 

\begin{figure}[t]
    \centering
    \begin{minipage}{0.47\textwidth}
        \centering
        \includegraphics[width=\textwidth]{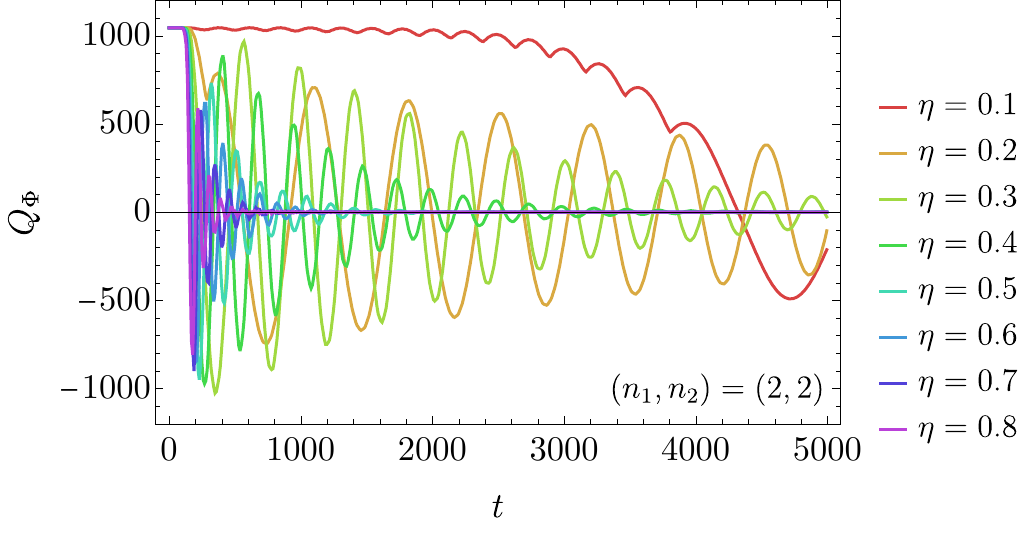}
    \end{minipage}
    \hspace{5mm}
    \begin{minipage}{0.47\textwidth}
        \centering
        \includegraphics[width=\textwidth]{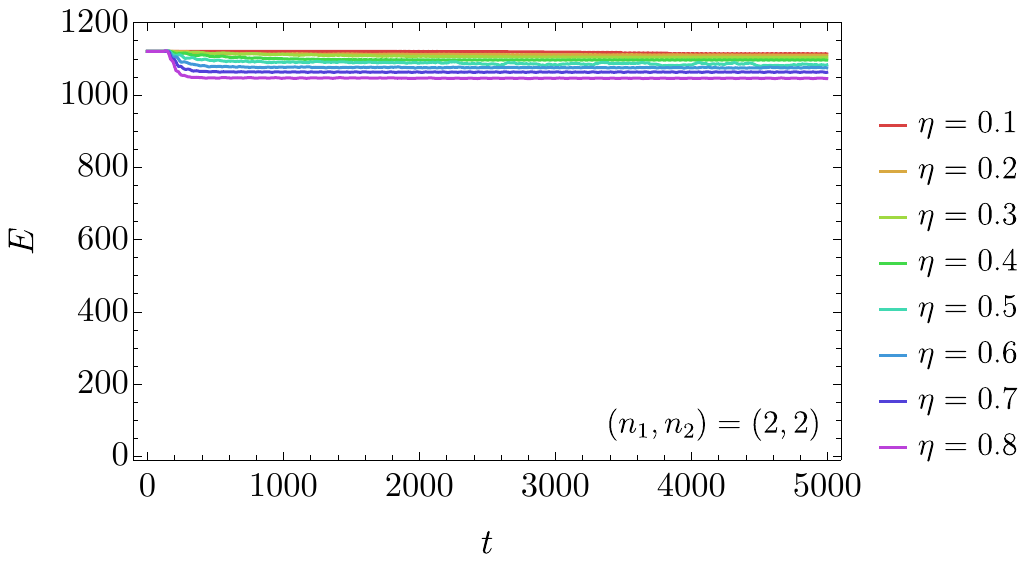}
    \end{minipage}
    \caption{%
        Same as the upper panels in Fig.~\ref{fig: n22 eta} except for fixing $|\Psi| = \eta$.
    }
    \label{fig: n22H eta}
\end{figure}

We show the results for $(n_1, n_2) = (2,2)$ in Fig.~\ref{fig: n22H eta}.
While the existence of stable configurations with small or vanishing charges is common with the results in the main text, the decrease in the charge and energy is milder.

\begin{figure}[t]
    \centering
    \begin{minipage}{0.47\textwidth}
        \centering
        \includegraphics[width=\textwidth]{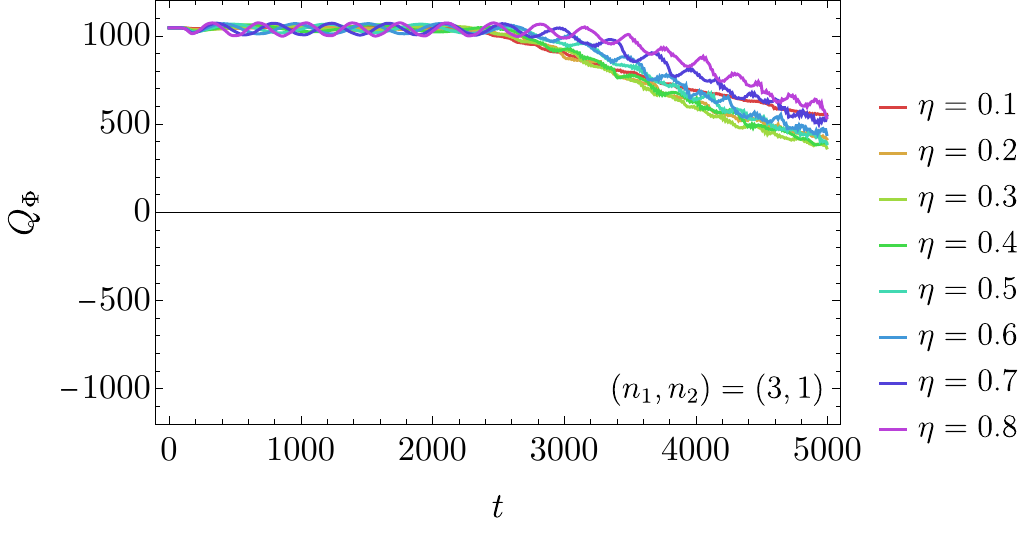}
    \end{minipage}
    \hspace{5mm}
    \begin{minipage}{0.47\textwidth}
        \centering
        \includegraphics[width=\textwidth]{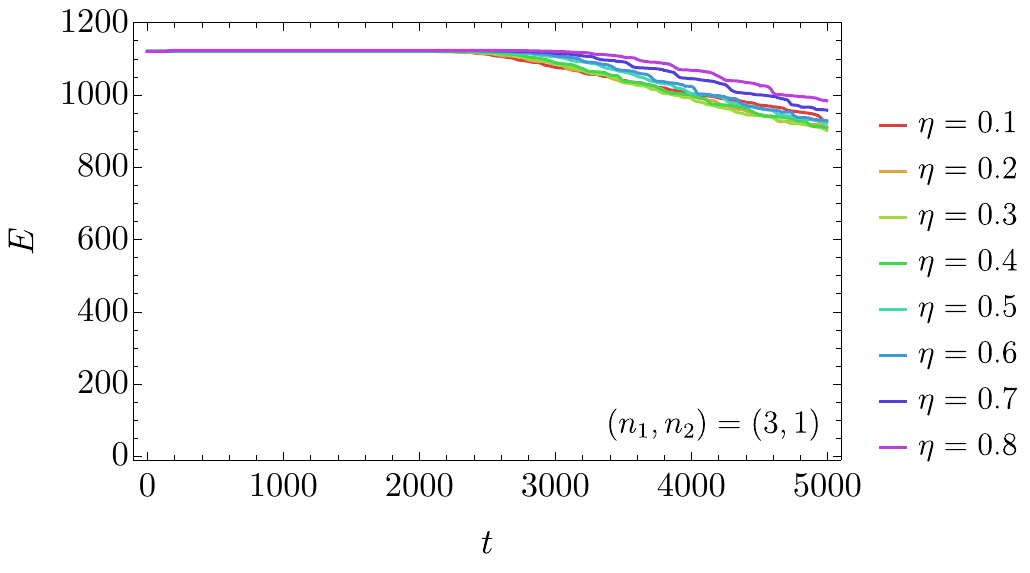}
    \end{minipage}
    \caption{%
        Same as the upper panels in Fig.~\ref{fig: n31 eta} except for fixing $|\Psi| = \eta$.
    }
    \label{fig: n31H eta}
\end{figure}

We show the results for $(n_1, n_2) = (3,1)$ in Fig.~\ref{fig: n31H eta}.
Compared with the results in the main text, the Q-balls start to decay later.
In addition, the decrease in the charge and energy is milder and has the opposite dependence on $\eta$.

\bibliographystyle{JHEP}
\bibliography{Ref}

\providecommand{\href}[2]{#2}\begingroup\raggedright\begin{thebibliography}{10}

\bibitem{Coleman:1985ki}
S.~R. Coleman, \emph{{Q-balls}},
  \href{https://doi.org/10.1016/0550-3213(86)90520-1}{\emph{Nucl. Phys. B}
  {\bfseries 262} (1985) 263}.

\bibitem{Banks:2010zn}
T.~Banks and N.~Seiberg, \emph{{Symmetries and Strings in Field Theory and
  Gravity}}, \href{https://doi.org/10.1103/PhysRevD.83.084019}{\emph{Phys. Rev.
  D} {\bfseries 83} (2011) 084019},
  [\href{https://arxiv.org/abs/1011.5120}{{\ttfamily 1011.5120}}].

\bibitem{Affleck:1984fy}
I.~Affleck and M.~Dine, \emph{{A New Mechanism for Baryogenesis}},
  \href{https://doi.org/10.1016/0550-3213(85)90021-5}{\emph{Nucl. Phys. B}
  {\bfseries 249} (1985) 361--380}.

\bibitem{Dine:1995kz}
M.~Dine, L.~Randall and S.~D. Thomas, \emph{{Baryogenesis from flat directions
  of the supersymmetric standard model}},
  \href{https://doi.org/10.1016/0550-3213(95)00538-2}{\emph{Nucl. Phys. B}
  {\bfseries 458} (1996) 291--326},
  [\href{https://arxiv.org/abs/hep-ph/9507453}{{\ttfamily hep-ph/9507453}}].

\bibitem{Kusenko:1997zq}
A.~Kusenko, \emph{{Solitons in the supersymmetric extensions of the standard
  model}}, \href{https://doi.org/10.1016/S0370-2693(97)00584-4}{\emph{Phys.
  Lett. B} {\bfseries 405} (1997) 108},
  [\href{https://arxiv.org/abs/hep-ph/9704273}{{\ttfamily hep-ph/9704273}}].

\bibitem{Kusenko:1997si}
A.~Kusenko and M.~E. Shaposhnikov, \emph{{Supersymmetric Q balls as dark
  matter}}, \href{https://doi.org/10.1016/S0370-2693(97)01375-0}{\emph{Phys.
  Lett. B} {\bfseries 418} (1998) 46--54},
  [\href{https://arxiv.org/abs/hep-ph/9709492}{{\ttfamily hep-ph/9709492}}].

\bibitem{Enqvist:1997si}
K.~Enqvist and J.~McDonald, \emph{{Q balls and baryogenesis in the MSSM}},
  \href{https://doi.org/10.1016/S0370-2693(98)00271-8}{\emph{Phys. Lett. B}
  {\bfseries 425} (1998) 309--321},
  [\href{https://arxiv.org/abs/hep-ph/9711514}{{\ttfamily hep-ph/9711514}}].

\bibitem{Enqvist:1998en}
K.~Enqvist and J.~McDonald, \emph{{B - ball baryogenesis and the baryon to dark
  matter ratio}},
  \href{https://doi.org/10.1016/S0550-3213(98)00695-6}{\emph{Nucl. Phys. B}
  {\bfseries 538} (1999) 321--350},
  [\href{https://arxiv.org/abs/hep-ph/9803380}{{\ttfamily hep-ph/9803380}}].

\bibitem{Kasuya:1999wu}
S.~Kasuya and M.~Kawasaki, \emph{{Q ball formation through Affleck-Dine
  mechanism}}, \href{https://doi.org/10.1103/PhysRevD.61.041301}{\emph{Phys.
  Rev. D} {\bfseries 61} (2000) 041301},
  [\href{https://arxiv.org/abs/hep-ph/9909509}{{\ttfamily hep-ph/9909509}}].

\bibitem{Kasuya:2000wx}
S.~Kasuya and M.~Kawasaki, \emph{{Q Ball formation in the gravity mediated SUSY
  breaking scenario}},
  \href{https://doi.org/10.1103/PhysRevD.62.023512}{\emph{Phys. Rev. D}
  {\bfseries 62} (2000) 023512},
  [\href{https://arxiv.org/abs/hep-ph/0002285}{{\ttfamily hep-ph/0002285}}].

\bibitem{Enqvist:2000gq}
K.~Enqvist, A.~Jokinen and J.~McDonald, \emph{{Flat direction condensate
  instabilities in the MSSM}},
  \href{https://doi.org/10.1016/S0370-2693(00)00578-5}{\emph{Phys. Lett. B}
  {\bfseries 483} (2000) 191--195},
  [\href{https://arxiv.org/abs/hep-ph/0004050}{{\ttfamily hep-ph/0004050}}].

\bibitem{Hiramatsu:2010dx}
T.~Hiramatsu, M.~Kawasaki and F.~Takahashi, \emph{{Numerical study of Q-ball
  formation in gravity mediation}},
  \href{https://doi.org/10.1088/1475-7516/2010/06/008}{\emph{JCAP} {\bfseries
  06} (2010) 008}, [\href{https://arxiv.org/abs/1003.1779}{{\ttfamily
  1003.1779}}].

\bibitem{Kawasaki:2005xc}
M.~Kawasaki, K.~Konya and F.~Takahashi, \emph{{Q-ball instability due to U(1)
  breaking}}, \href{https://doi.org/10.1016/j.physletb.2005.05.082}{\emph{Phys.
  Lett. B} {\bfseries 619} (2005) 233--239},
  [\href{https://arxiv.org/abs/hep-ph/0504105}{{\ttfamily hep-ph/0504105}}].

\bibitem{Kasuya:2014ofa}
S.~Kasuya and M.~Kawasaki, \emph{{Baryogenesis from the gauge-mediation type
  Q-ball and the new type of Q-ball as the dark matter}},
  \href{https://doi.org/10.1103/PhysRevD.89.103534}{\emph{Phys. Rev. D}
  {\bfseries 89} (2014) 103534},
  [\href{https://arxiv.org/abs/1402.4546}{{\ttfamily 1402.4546}}].

\bibitem{Cotner:2016dhw}
E.~Cotner and A.~Kusenko, \emph{{Astrophysical constraints on dark-matter
  $Q$-balls in the presence of baryon-violating operators}},
  \href{https://doi.org/10.1103/PhysRevD.94.123006}{\emph{Phys. Rev. D}
  {\bfseries 94} (2016) 123006},
  [\href{https://arxiv.org/abs/1609.00970}{{\ttfamily 1609.00970}}].

\bibitem{Kawasaki:2019ywz}
M.~Kawasaki and H.~Nakatsuka, \emph{{Q-ball decay through A-term in the
  gauge-mediated SUSY breaking scenario}},
  \href{https://doi.org/10.1088/1475-7516/2020/04/017}{\emph{JCAP} {\bfseries
  04} (2020) 017}, [\href{https://arxiv.org/abs/1912.06993}{{\ttfamily
  1912.06993}}].

\bibitem{Dine:2003ax}
M.~Dine and A.~Kusenko, \emph{{The Origin of the matter - antimatter
  asymmetry}}, \href{https://doi.org/10.1103/RevModPhys.76.1}{\emph{Rev. Mod.
  Phys.} {\bfseries 76} (2003) 1},
  [\href{https://arxiv.org/abs/hep-ph/0303065}{{\ttfamily hep-ph/0303065}}].

\bibitem{Enqvist:2003gh}
K.~Enqvist and A.~Mazumdar, \emph{{Cosmological consequences of MSSM flat
  directions}},
  \href{https://doi.org/10.1016/S0370-1573(03)00119-4}{\emph{Phys. Rept.}
  {\bfseries 380} (2003) 99--234},
  [\href{https://arxiv.org/abs/hep-ph/0209244}{{\ttfamily hep-ph/0209244}}].

\bibitem{Kusenko:1997it}
A.~Kusenko, M.~E. Shaposhnikov, P.~G. Tinyakov and I.~I. Tkachev, \emph{{Star
  wreck}}, \href{https://doi.org/10.1016/S0370-2693(98)00133-6}{\emph{Phys.
  Lett. B} {\bfseries 423} (1998) 104--108},
  [\href{https://arxiv.org/abs/hep-ph/9801212}{{\ttfamily hep-ph/9801212}}].

\bibitem{Kusenko:2005du}
A.~Kusenko, L.~C. Loveridge and M.~Shaposhnikov, \emph{{Astrophysical bounds on
  supersymmetric dark-matter Q-balls}},
  \href{https://doi.org/10.1088/1475-7516/2005/08/011}{\emph{JCAP} {\bfseries
  08} (2005) 011}, [\href{https://arxiv.org/abs/astro-ph/0507225}{{\ttfamily
  astro-ph/0507225}}].

\bibitem{Peccei:1977hh}
R.~D. Peccei and H.~R. Quinn, \emph{{CP Conservation in the Presence of
  Instantons}}, \href{https://doi.org/10.1103/PhysRevLett.38.1440}{\emph{Phys.
  Rev. Lett.} {\bfseries 38} (1977) 1440--1443}.

\bibitem{Peccei:1977ur}
R.~D. Peccei and H.~R. Quinn, \emph{{Constraints Imposed by CP Conservation in
  the Presence of Instantons}},
  \href{https://doi.org/10.1103/PhysRevD.16.1791}{\emph{Phys. Rev. D}
  {\bfseries 16} (1977) 1791--1797}.

\bibitem{Preskill:1982cy}
J.~Preskill, M.~B. Wise and F.~Wilczek, \emph{{Cosmology of the Invisible
  Axion}}, \href{https://doi.org/10.1016/0370-2693(83)90637-8}{\emph{Phys.
  Lett. B} {\bfseries 120} (1983) 127--132}.

\bibitem{Abbott:1982af}
L.~F. Abbott and P.~Sikivie, \emph{{A Cosmological Bound on the Invisible
  Axion}}, \href{https://doi.org/10.1016/0370-2693(83)90638-X}{\emph{Phys.
  Lett. B} {\bfseries 120} (1983) 133--136}.

\bibitem{Dine:1982ah}
M.~Dine and W.~Fischler, \emph{{The Not So Harmless Axion}},
  \href{https://doi.org/10.1016/0370-2693(83)90639-1}{\emph{Phys. Lett. B}
  {\bfseries 120} (1983) 137--141}.

\bibitem{Kobayashi:2025qao}
J.~Kobayashi, K.~Nakayama and M.~Yamada, \emph{{Decay rate of PQ-ball}},
  \href{https://arxiv.org/abs/2504.13510}{{\ttfamily 2504.13510}}.

\bibitem{Bogolyubsky:1976yu}
I.~L. Bogolyubsky and V.~G. Makhankov, \emph{{Lifetime of Pulsating Solitons in
  Some Classical Models}}, {\emph{Pisma Zh. Eksp. Teor. Fiz.} {\bfseries 24}
  (1976) 15--18}.

\bibitem{Gleiser:1993pt}
M.~Gleiser, \emph{{Pseudostable bubbles}},
  \href{https://doi.org/10.1103/PhysRevD.49.2978}{\emph{Phys. Rev. D}
  {\bfseries 49} (1994) 2978--2981},
  [\href{https://arxiv.org/abs/hep-ph/9308279}{{\ttfamily hep-ph/9308279}}].

\bibitem{Kasuya:2002zs}
S.~Kasuya, M.~Kawasaki and F.~Takahashi, \emph{{I-balls}},
  \href{https://doi.org/10.1016/S0370-2693(03)00344-7}{\emph{Phys. Lett. B}
  {\bfseries 559} (2003) 99--106},
  [\href{https://arxiv.org/abs/hep-ph/0209358}{{\ttfamily hep-ph/0209358}}].

\bibitem{Kawasaki:2025}
M.~Kawasaki, K.~Murai and F.~Takahashi. In preparation, 2025.

\bibitem{Salmi:2012ta}
P.~Salmi and M.~Hindmarsh, \emph{{Radiation and Relaxation of Oscillons}},
  \href{https://doi.org/10.1103/PhysRevD.85.085033}{\emph{Phys. Rev. D}
  {\bfseries 85} (2012) 085033},
  [\href{https://arxiv.org/abs/1201.1934}{{\ttfamily 1201.1934}}].

\bibitem{Kawasaki:2015vga}
M.~Kawasaki, F.~Takahashi and N.~Takeda, \emph{{Adiabatic Invariance of
  Oscillons/I-balls}},
  \href{https://doi.org/10.1103/PhysRevD.92.105024}{\emph{Phys. Rev. D}
  {\bfseries 92} (2015) 105024},
  [\href{https://arxiv.org/abs/1508.01028}{{\ttfamily 1508.01028}}].

\bibitem{Mukaida:2016hwd}
K.~Mukaida, M.~Takimoto and M.~Yamada, \emph{{On Longevity of
  I-ball/Oscillon}}, \href{https://doi.org/10.1007/JHEP03(2017)122}{\emph{JHEP}
  {\bfseries 03} (2017) 122},
  [\href{https://arxiv.org/abs/1612.07750}{{\ttfamily 1612.07750}}].

\bibitem{Ibe:2019vyo}
M.~Ibe, M.~Kawasaki, W.~Nakano and E.~Sonomoto, \emph{{Decay of I-ball/Oscillon
  in Classical Field Theory}},
  \href{https://doi.org/10.1007/JHEP04(2019)030}{\emph{JHEP} {\bfseries 04}
  (2019) 030}, [\href{https://arxiv.org/abs/1901.06130}{{\ttfamily
  1901.06130}}].

\bibitem{Ibe:2019lzv}
M.~Ibe, M.~Kawasaki, W.~Nakano and E.~Sonomoto, \emph{{Fragileness of Exact
  I-ball/Oscillon}},
  \href{https://doi.org/10.1103/PhysRevD.100.125021}{\emph{Phys. Rev. D}
  {\bfseries 100} (2019) 125021},
  [\href{https://arxiv.org/abs/1908.11103}{{\ttfamily 1908.11103}}].

\bibitem{Zhang:2020bec}
H.-Y. Zhang, M.~A. Amin, E.~J. Copeland, P.~M. Saffin and K.~D. Lozanov,
  \emph{{Classical Decay Rates of Oscillons}},
  \href{https://doi.org/10.1088/1475-7516/2020/07/055}{\emph{JCAP} {\bfseries
  07} (2020) 055}, [\href{https://arxiv.org/abs/2004.01202}{{\ttfamily
  2004.01202}}].

\bibitem{Fedderke:2025sic}
M.~A. Fedderke, J.~Huang and N.~Siemonsen, \emph{{Periodic Cosmic String
  Formation and Dynamics}},  \href{https://arxiv.org/abs/2503.03116}{{\ttfamily
  2503.03116}}.

\bibitem{Kasuya:1999wv}
S.~Kasuya, \emph{{Q-ball formation and its properties}},  in \emph{{7th
  International Symposium on Particles, Strings and Cosmology}}, pp.~301--304,
  12, 1999.

\bibitem{Enqvist:2002si}
K.~Enqvist, S.~Kasuya and A.~Mazumdar, \emph{{Inflatonic solitons in running
  mass inflation}},
  \href{https://doi.org/10.1103/PhysRevD.66.043505}{\emph{Phys. Rev. D}
  {\bfseries 66} (2002) 043505},
  [\href{https://arxiv.org/abs/hep-ph/0206272}{{\ttfamily hep-ph/0206272}}].

\end{thebibliography}\endgroup

\end{document}